\documentclass[superscriptaddress,twocolumn,pre]{revtex4}

\usepackage{amsmath}
\usepackage{graphicx,color}

\newcommand{\be}{\begin{equation}}
\newcommand{\ee}{\end{equation}}
\newcommand{\bea}{\begin{eqnarray}}
\newcommand{\eea}{\end{eqnarray}}

\begin{document}
\sloppy

%-title page-%

\title{Exact analytical solution of the collapse of self-gravitating \\
Brownian  particles and bacterial populations at zero temperature}

\author{Pierre-Henri Chavanis}
%\email{chavanis@irsamc.ups-tlse.fr}
\affiliation{Laboratoire de Physique Th\'eorique (IRSAMC), CNRS and UPS, Universit\'e Paul Sabatier, F-31062 Toulouse, France}
\author{Cl\'ement Sire}
%\email{clement.sire@irsamc.ups-tlse.fr}
\affiliation{Laboratoire de Physique Th\'eorique (IRSAMC), CNRS and UPS, Universit\'e Paul Sabatier, F-31062 Toulouse, France}

\begin{abstract}
We provide an exact analytical solution of the collapse dynamics of
self-gravitating Brownian particles and bacterial populations at zero
temperature. These systems are described by the Smoluchowski-Poisson
system or Keller-Segel model in which the diffusion term is
neglected. As a result, the dynamics is purely deterministic. A cold
system undergoes a gravitational collapse leading to a finite time
singularity: the central density increases and becomes infinite in a
finite time $t_{coll}$. The evolution continues in the post collapse
regime. A Dirac peak emerges, grows and finally captures all the mass
in a finite time $t_{end}$, while the central density excluding the
Dirac peak progressively decreases. Close to the collapse time, the
pre and post collapse evolutions are self-similar. Interestingly, if one
starts from a parabolic density profile, one obtains an exact
analytical solution that describes the whole collapse dynamics, from
the initial time to the end, and accounts for non self-similar
corrections that were neglected in previous works. Our results have
possible application in different areas including astrophysics,
chemotaxis, colloids and nanoscience.
\end{abstract}

\maketitle

%-section: Introduction-%

\section{Introduction}

In a series of papers (see \cite{grossmann} for a short review), we
have studied a class of mean field drift-diffusion equations of the
form
\begin{equation}
\label{i1}
\frac{\partial\rho}{\partial t}=\nabla\cdot\left \lbrack\frac{1}{\xi}\left ( \nabla p+\rho\nabla\Phi\right )\right\rbrack,
\end{equation}
\begin{equation}
\label{i2}
\epsilon\frac{\partial\Phi}{\partial t}=\Delta\Phi-k^2\Phi-S_d G\rho,
\end{equation}
where the equation of state $p=p(\rho)$ can take different
shapes. These equations, derived in a statistical mechanics context in
\cite{gen,nfp}, can be viewed as nonlinear mean field Fokker-Planck
(NFP) equations \cite{frank} describing a system of Langevin particles
in interaction \footnote{We can obtain a larger class of models (see
\cite{nfp} and Appendix
\ref{sec_nfp}) by letting the friction  coefficient $\xi$ depend on the density, $\xi=\xi(\rho)$, or more generally be a function of position and time,
$\xi=\xi({\bf r},t)$.}. The first equation (\ref{i1}) can be
interpreted as a generalized Smoluchowski equation in which the
evolution of the density $\rho({\bf r},t)$ is due to a competition
between a (nonlinear) diffusion and a drift. Contrary to the ordinary
Smoluchowski equation \cite{risken}, the potential $\Phi({\bf r},t)$
is not fixed but determined by the density itself according to the
reaction-diffusion equation (\ref{i2}). Equations
(\ref{i1})-(\ref{i2}) are based on a mean field approximation which is
known to be exact for long-range interactions when the number of
particles $N\rightarrow +\infty$ \cite{cdr,hb2}. On the other hand, the
linear diffusion $\nabla \cdot (D\nabla\rho)$ in the ordinary
Smoluchowski equation is replaced by a more general term $\nabla \cdot
(\xi^{-1}\nabla p)$ which can lead to anomalous diffusion. The
function $p({\bf r},t)=p\lbrack \rho({\bf r},t)\rbrack$ can be
interpreted as a barotropic pressure \cite{nfp}. It can take into
account microscopic constraints such as short-range interactions,
close packing effects, steric hindrance, non-extensivity, exclusion or
inclusion principles... Accordingly, the drift-diffusion equations
(\ref{i1})-(\ref{i2}) are associated with generalized forms of free
energy that can be non-Boltzmannian. This leads to a notion of
effective generalized thermodynamics (EGT) \cite{nfp,frank}.

The drift-diffusion equations (\ref{i1})-(\ref{i2}) can be derived
heuristically from generalized stochastic processes in physical space
of the form
\begin{equation}
\label{i3}
\frac{d{\bf r}}{dt}=-\frac{1}{\xi}\nabla\Phi+\sqrt{\frac{2p(\rho)}{\xi\rho}}{\bf R}(t),
\end{equation}
where $\Phi({\bf r},t)$ is a mean field potential determined by
Eq. (\ref{i2}) and ${\bf R}(t)$ is a Gaussian white noise satisfying
$\langle {\bf R}(t)\rangle={\bf 0}$ and $\langle
{R}_i(t)R_j(t')\rangle=\delta_{ij}\delta(t-t')$ where $i=1,...,d$
label the coordinates of space. The Fokker-Planck equation associated
with the generalized Langevin equation (\ref{i3}) is the generalized
mean field Smoluchowski equation (\ref{i1}). Note that the strength of
the noise can depend on the local density $\rho({\bf r},t)$ of
particles which may lead to anomalous diffusion \cite{borland}. The
generalized mean field Smoluchowski equation (\ref{i1}) can also be
derived from the master equation by assuming that the probabilities of
transition explicitly depend on the occupation numbers of the initial
and arrival states (see \cite{kaniadakis} and Sec. 2.11 of
\cite{nfp}). This kinetical interaction principle (KIP) takes into
account microscopic constraints such as exclusion or inclusion
principles \cite{kaniadakis}.

The model (\ref{i1})-(\ref{i2}) assumes an overdamped evolution in
which the velocity of the particles is directly proportional to the
force $-\nabla\Phi$ as in Eq. (\ref{i3}), but more general models
taking into account inertial effects can be introduced
\cite{nfp}. They are based on generalized stochastic processes in
phase space of the form
\begin{equation}
\label{i4a}
\frac{d{\bf r}}{dt}={\bf v},
\end{equation}
\begin{equation}
\label{i4}
\frac{d{\bf v}}{dt}=-\xi{\bf v}-\nabla\Phi+\sqrt{2Df\left\lbrack\frac{C(f)}{f}\right\rbrack'}{\bf R}(t),
\end{equation}
where the noise depends on the distribution function $f({\bf r},{\bf v},t)$
of the particles. Here, $\xi$ is a friction coefficient and
$C(f)$ is a convex function ($C''>0$). This function determines a
generalized entropic functional $S=-\int C(f)\, d{\bf r}d{\bf v}$
\cite{nfp}. The Fokker-Planck equation corresponding to the
generalized Langevin equation (\ref{i4}) is the generalized mean field
Kramers equation
\begin{equation}
\label{i5}
\frac{\partial f}{\partial t}+{\bf v}\cdot\frac{\partial f}{\partial {\bf r}}-\nabla\Phi\cdot \frac{\partial f}{\partial {\bf v}}=\frac{\partial}{\partial {\bf v}}\cdot\left\lbrack \xi\left (T C''(f) f \frac{\partial f}{\partial {\bf v}}+ f {\bf v}\right )\right\rbrack,
\end{equation}
\begin{equation}
\label{i6}
\epsilon\frac{\partial\Phi}{\partial t}=\Delta\Phi-k^2\Phi-S_d G\rho,
\end{equation}
where we have defined the generalized thermodynamical temperature $T$
through an Einstein-like formula $D=\xi T$ \cite{nfp}. These equations
govern the evolution of the distribution function $f({\bf r},{\bf
v},t)$ in phase space. The generalized Smoluchowski equation (\ref{i1}) can be
derived from the generalized Kramers equation (\ref{i5}) in a strong
friction limit $\xi\rightarrow +\infty$ by using a Chapman-Enskog
expansion \cite{cll} or a method of moments \cite{nfp}. In the strong
friction limit, there exists a precise link \cite{nfp} between the
barotropic pressure $p(\rho)$ occuring in the generalized Smoluchowski
equation (\ref{i1}) and the function $C(f)$ occurring in the
generalized Kramers equation (\ref{i5}). For example, the Boltzmann
entropy $S_B=-\int \frac{f}{m}\ln \frac{f}{m}\, d{\bf r}d{\bf v}$
(leading to normal diffusion) \cite{risken} determines an isothermal
equation of state $p({\bf r},t)=\rho({\bf r},t)k_B T/m$, where $T$ is
the temperature and $m$ the individual mass of the particles. The
Tsallis entropy $S_q=-\frac{1}{q-1}\int (f^q-f)\, d{\bf r}d{\bf v}$
(leading to anomalous algebraic diffusion)
\cite{tsallis} determines a polytropic equation of state $p({\bf
r},t)=K\rho({\bf r},t)^{\gamma}$, where $K$ is the polytropic constant
and $\gamma$ the polytropic index. They can be related to $T$ and $q$
as explained in \cite{nfp}. We can also consider Fermi-Dirac and
Bose-Einstein entropies (taking into account exclusion or inclusion
constraints) determining fermionic and bosonic equations of state
\cite{nfp}. For example, the equation of state
$p(\rho)=-T\sigma_0\ln(1-\rho/\sigma_0)$ that takes into account an
exclusion constraint in physical space has been studied in
\cite{degrad}.

The generalized Smoluchowski equation (\ref{i1}) can also be derived from the damped Euler equations \cite{gen,nfp}:
\begin{equation}
\label{der1}
\frac{\partial\rho}{\partial t}+\nabla\cdot\left (\rho {\bf u}\right )=0,
\end{equation}
\begin{equation}
\label{der2}
\frac{\partial {\bf u}}{\partial t}+({\bf u}\cdot \nabla){\bf u}=-\frac{1}{\rho}\nabla p-\nabla\Phi-\xi {\bf u}.
\end{equation}
The first equation is the equation of continuity (taking into account the local conservation of mass) and the second equation is the momentum equation. We have assumed that the pressure is isotropic and barotropic, i.e. it is a function of the density $p=p(\rho)$. In the ideal case, it is given by the perfect gas law $p({\bf r},t)=\rho({\bf r},t) k_B T/m$. More generally, we allow $p(\rho)$ to be nonlinear so as to  take into account microscopic constraints such as short-range interactions and other non ideal effects. In the strong friction limit $\xi\rightarrow +\infty$, we can formally ignore the inertia of the particles in the damped Euler equation leading to the generalized Darcy law
\begin{equation}
\label{der2b}
\xi {\bf u}\simeq -\frac{1}{\rho}\nabla p-\nabla\Phi.
\end{equation}
Substituting this relation in the continuity equation (\ref{der1}), we
obtain the generalized Smoluchowski equation (\ref{i1}). We note,
however, that the damped Euler equations (\ref{der1})-(\ref{der2})
cannot be directly derived from the stochastic equations
(\ref{i4a})-(\ref{i4}) or from the Kramers equation (\ref{i5}) as they
rely on a {\it local thermodynamic equilibrium} (LTE) assumption \cite{gen}
that is not rigorously justified. By contrast, the generalized
Smoluchowski equation (\ref{i1}) can be rigorously derived from the
generalized Kramers equation (\ref{i5}) in a strong friction limit
$\xi\rightarrow +\infty$ as shown in Refs \cite{nfp,cll}.

We finally note that the generalized mean field Smoluchowski equation
(\ref{i1})-(\ref{i2}) has recently been derived from a kinetic theory
\cite{longshort} that combines the technics used in the theory of
systems with long-range interactions and in the theory of liquids (see
Appendix \ref{sec_deriv}). In this approach, the long-range
interactions (leading to the drift) are modeled by using the mean
field theory and the short-range interactions (leading to the pressure
term) are modeled by using the dynamic density functional theory
(DDFT). This new derivation
\cite{longshort} provides an alternative to the kinetic theory based
on generalized thermodynamics
\cite{nfp,frank}.

The generalized mean field drift-diffusion model (\ref{i1})-(\ref{i2})
appears in a number of physical situations that we briefly review.

(i) When $\Phi=\Phi_{ext}({\bf r})$ is an external potential and
$p=\rho k_B T/m$, we recover the ordinary Smoluchowski equation
describing, for example, the sedimentation of colloidal particles in a
gravitational field \cite{risken}. When $\Phi_{ext}=0$ and
$p=K\rho^{\gamma}$, we recover the porous medium equation
\cite{spohn}. The porous medium equation with an external potential
$\Phi_{ext}({\bf r})$ has been considered by Plastino \& Plastino \cite{pp} in
connection with Tsallis generalized thermodynamics \cite{tsallis}.

(ii) When $\epsilon=k=0$, and for an isothermal equation of state
$p=\rho k_B T/m$, we get the Smoluchowski-Poisson system.  In the
repulsive case $G<0$, i.e. for the Coulombian interaction, this
corresponds to the Debye-H\"uckel model of electrolytes \cite{dh}. The
same equations, called the Nernst-Planck equations \cite{biler}, are
also used to study ion transport in biological channels
\cite{barcilon,syganow} and carrier transport in semiconductors
\cite{semi}. In the attractive case $G>0$, i.e. for the gravitational
interaction, the Smoluchowski-Poisson system describes a gas of
self-gravitating Brownian particles \cite{grossmann}. In that case,
$G$ represents the constant of gravity and $S_d$ is the surface of a
unit sphere in $d$ dimensions \footnote{To avoid misunderstandings, we
stress that the Smoluchowski-Poisson system does {\it not} describe
traditional astrophysical systems. Indeed, astrophysical systems are
not in a strong friction limit but rather in a weak (or no) friction
limit. For example, stars are described by the Euler-Poisson system
(hydrodynamics) \cite{chandra}, galaxies by the Vlasov-Poisson system
(collisionless dynamics) \cite{bt} and globular clusters by the
Vlasov-Landau-Poisson system (collisional dynamics)
\cite{spitzer}. The Smoluchowski-Poisson system could describe the
dynamics of dust particles in the protoplanetary nebula, where
particles experience a friction with the gas and a stochastic force
due to turbulence or other diffusive effects \cite{aa}. Unfortunately,
when the dust particles are small, the friction force is strong but
the gravitational interaction is weak. Alternatively, when the
particles are large, the gravitational interaction is strong but the
friction force is weak. Therefore, the Smoluchowski-Poisson system is
only valid in an intermediate regime which may not be the most
relevant. Nevertheless, in the context of the statistical mechanics of
self-gravitating systems
\cite{paddy,ijmpb}, the model of self-gravitating Brownian particles
is interesting at a conceptual level because it provides a dynamical
model associated with the {\it canonical ensemble} \cite{grossmann}
while stellar systems are associated with the {\it microcanonical
ensemble}
\cite{antonov,lbw}. It can be used therefore to illustrate the notion
of ensemble inequivalence that is generic for systems with long-range
interactions \cite{cdr}.}.

(iii) When  $\epsilon\ge 0$ and $k\ge 0$, we obtain a generalized form of Keller-Segel model \cite{ks} describing the chemotaxis of bacterial populations. In that case, $\rho$ denotes the density of bacteria and $c=-\Phi$ the concentration of the secreted chemical. Furthermore, the term $-k^2\Phi$ takes into account a possible degradation of the chemical, leading to a shielding of the interaction on a scale $k^{-1}$ \cite{degrad}.

(iv) Recently, it has been shown \cite{colloids} that Eqs. (\ref{i1})-(\ref{i2}) with $\epsilon=0$, $k\ge 0$ and $d=2$, describe the dynamics of colloids at a fluid interface driven by attractive capillary interactions. In that context, $\rho$ is the particle density, $\Phi$ the interfacial deformation and $k^{-1}$ the capillary length whose finiteness leads to a shielding of the interaction. Finally, $G$ is equal to the ratio between the capillary monopole associated to a single particle and the surface tension.

(v) Drift-diffusion equations of the form (\ref{i1})-(\ref{i2}), or their generalization (see Appendix \ref{sec_nfp}), have been proposed to describe phase segregation in model alloys with long-range interactions \cite{gl} and aggregation of finite-size particles and directed-assembly processes for nanoscience \cite{holm}.

(vi) Nonlinear mean field Fokker-Planck equations of the form  (\ref{i1})-(\ref{i2}) and (\ref{i5})-(\ref{i2}) have also been introduced in two-dimensional (2D) vortex dynamics and stellar dynamics in order to provide  a small-scale parametrization of the 2D Euler-Poisson and Vlasov-Poisson systems \cite{rs,csr}. They describe the rapid formation of quasi stationary states (QSS) on the coarse-grained scale representing large scale vortices or galaxies \cite{houches}. In 2D hydrodynamics, $\rho$ corresponds to the coarse-grained vorticity $\overline{\omega}$ and $\Phi$ to the stream function $\psi$. On the other hand, Eq. (\ref{i2}) reduces to the Poisson equation $-\Delta\psi=\overline{\omega}$. In the case of geophysical flows described by the quasi geostrophic (QG) equations, the Poisson equation is replaced by a screened Poisson equation $-\Delta\psi+\psi/R^2=\overline{q}$ where $R$ is the Rossby radius and $\overline{q}$ the coarse-grained potential vorticity. In stellar dynamics, $\overline{f}$ is the coarse-grained distribution function, $\rho$ is the density of stars and $\Phi$ the gravitational potential.

The numerous analogies between these very different systems, covering
all scales of physics (self-gravitating systems, large-scale vortices,
biological organisms, electrolytes, colloids, semiconductors,
nanotechnology,...) show the importance of studying nonlinear mean
field Fokker-Planck equations of the form (\ref{i1})-(\ref{i2}) and
(\ref{i5})-(\ref{i2}) \footnote{The theoretical interest of these
equations was stressed early in \cite{gen} before all their possible
applications were realized.}.  Depending on the equation of state
$p(\rho,T)$, on the value of the temperature $T$ and on the dimension of
space $d$, the system can either (i) converge towards an equilibrium
state, (ii) collapse, or (iii) evaporate. A short description of these
different regimes in the case of equations (\ref{i1})-(\ref{i2}) with
$\epsilon=k=0$ and in the case of isothermal and polytropic equations
of state has been given in \cite{grossmann}.  It reveals the
complexity and the richness of these apparently simple equations.

In this paper, we treat in detail a simple case corresponding to
$\epsilon=k=0$, $G>0$ (long-range Newtonian interaction) and $p=0$
(cold system) \footnote{The case of repulsive interactions $G<0$ is
treated in Appendix \ref{sec_repu}.}. In other words, we consider the
overdamped dynamics of particles in Newtonian interaction in the zero
temperature limit $T=0$. This system undergoes a gravitational
collapse resulting ultimately in the formation of a Dirac peak
containing all the mass. As discussed above, this result can have
application in other areas such as chemotaxis, colloid dynamics and
nanotechnology. Interestingly, if we start from a parabolic density
profile, we can obtain an exact analytical solution of the equations
for all times. This solution describes both pre and post collapse
regimes. The pre-collapse regime leads to the formation of a finite
time singularity: the central density increases like
$(t_{coll}-t)^{-1}$ and becomes infinite in a finite time $t_{coll}$
resulting in a singular density profile proportional to
$r^{-{2d/}{(d+2)}}$. The evolution continues in the post-collapse
regime with the formation of a Dirac peak at $r=0$. The mass contained
in the Dirac peak grows like
$(t_{coll}/t)^{(d+2)/2}(t-t_{coll})^{d/2}$ while the central density
excluding the Dirac peak decreases like
$(t-t_{coll})^{-1}(t_{end}/t-1)$. Finally, in a finite time
$t_{end}=((d+2)/2) t_{coll}$, all the particles are ultimately
absorbed by the Dirac peak. In our previous works \cite{crs,sc,post},
we had only described the self-similar dynamics of the system at $T=0$
close to the collapse time $t_{coll}$ and for $r\rightarrow 0$. The
present analytical solution extends our results to all times $0\le
t\le t_{end}$ and all radii $0\le r\le r_{max}(t)$, and describes
exact corrections to the self-similar solution. It provides a simple
illustration of the Dirac peak formation in the post-collapse
regime. These behaviors (pre and post collapse, finite time
singularity, growth of a Dirac peak) also arise at $T\neq 0$ (with
different exponents) but they are more difficult to investigate
analytically \cite{crs,sc,post,lushnikov}. Therefore, the present
explicit solution can be useful before considering more complicated
situations.

\section{The formal solution of the problem}
\label{sec_formal}

The Smoluchowski-Poisson (SP) system at $T=0$ reduces to the form
\begin{equation}
\label{f1}
\frac{\partial\rho}{\partial t}=\nabla\cdot\left (\frac{1}{\xi}\rho\nabla\Phi\right ),
\end{equation}
\begin{equation}
\label{f2}
\Delta\Phi=S_d G\rho.
\end{equation}
Equation (\ref{f1}) is equivalent to a continuity equation
\begin{equation}
\label{f3}
\frac{\partial\rho}{\partial t}+\nabla\cdot (\rho {\bf u})=0,
\end{equation}
with a velocity field
\begin{equation}
\label{f4}
{\bf u}({\bf r},t)=-\frac{1}{\xi}\nabla\Phi({\bf r},t).
\end{equation}
Introducing a Lagrangian description of motion like in \cite{crs}, the equation determining the trajectory of a ``fluid'' particle is
\begin{equation}
\label{f5}
\frac{d{\bf r}}{dt}={\bf u}({\bf r},t)=-\frac{1}{\xi}\nabla\Phi({\bf r},t).
\end{equation}
It coincides with the equation of motion (\ref{i3}) when the noise is switched off ($T=0$). In that case, the dynamics is deterministic. If the distribution of particles is initially spherically symmetric, it will remain spherically symmetric for all times. Using the Gauss theorem, the equation of motion (\ref{f5}) becomes
\begin{equation}
\label{f6}
\frac{d{r}}{dt}=-\frac{1}{\xi}\frac{GM(r,t)}{r^{d-1}},
\end{equation}
where
\begin{equation}
\label{f7}
M(r,t)=\int_0^r \rho(r_1,t)S_d r_1^{d-1}dr_1,
\end{equation}
is the mass contained within the sphere of radius $r$
at time $t$. Conversely,
\begin{equation}
\label{f8}
\rho(r,t)=\frac{1}{S_d r^{d-1}}\frac{\partial M}{\partial r}(r,t).
\end{equation}
Since the equation of motion (\ref{f6}) is a first order differential equation, the particles do not cross each other. Therefore, the conservation of mass implies
\begin{equation}
\label{f9}
M(r,t)=M(a,0),
\end{equation}
where $r$ is the position at time $t$ of the particle located at $r=a$ at $t=0$. Equation (\ref{f6}) can therefore be rewritten
\begin{equation}
\label{f10}
\frac{d{r}}{dt}=-\frac{1}{\xi}\frac{GM(a,0)}{r^{d-1}},
\end{equation}
and it is easily integrated into
\begin{equation}
\label{f11}
r^d=a^d-\frac{d}{\xi}GM(a,0)t.
\end{equation}
Equations (\ref{f9}) and (\ref{f11}) give the general exact solution of the problem \cite{crs}. The time at which the particle initially located at $r=a$ reaches the origin $r=0$ is
\begin{equation}
\label{f11b}
t_*(a)=\frac{\xi a^d}{dGM(a,0)}.
\end{equation}
This time increases with the distance provided that $M(a,0)/a^d$ is a decreasing function of $a$ (physical case). The opposite situation is treated in Appendix \ref{sec_peri}.

{\it Remark:} the gravitational collapse of a cold gas ($T=0$)
initially at rest described by the Euler-Poisson system has been
investigated long ago by Hunter \cite{hunter}, Mestel \cite{mestel}
and Penston \cite{penston} in astrophysics. In that case,
Eq. (\ref{f10}) is replaced by a second order differential equation
which can be solved in a parametric form. The collapse of a stellar
system which starts from a configuration in which all the stars have
zero velocity is initially similar to that of a cold gas at
$T=0$. However, due to fluctuations (finite $N$ effects), the velocity
dispersion of the stars increases and orbit crossing occurs \cite{bt}.
In that case, the mass interior to a given particle changes so that
the collapse solution becomes much more complicated. The system
undergoes damped oscillations before finally settling on a virialized
state as a result of violent relaxation \cite{lb}. The simplification
obtained by considering self-gravitating Brownian particles in the
overdamped limit allows us to obtain a complete and explicit solution
of the collapse dynamics for all times since there is no orbit
crossing in that case. As we shall see, the system creates a Dirac peak
at the origin instead of exhibiting damped oscillations.

\section{The collapse of a cold homogeneous sphere}
\label{sec_hom}

Let us first consider the collapse of a homogeneous sphere of mass $M$ and initial radius $R$. Its initial density is $\rho(a,0)=dM/(S_d R^d)$ for $a\le R$ and $\rho(a,0)=0$ for $a\ge R$. The corresponding mass profile is
\begin{equation}
\label{h1}
M(a,0)=\frac{M}{R^d}a^d,
\end{equation}
for $a\le R$ and $M(a,0)=M$ for $a\ge R$. According to Eqs. (\ref{f11})
and (\ref{h1}), the position at time $t$ of a particle initially located at
$r=a$ is
\begin{equation}
\label{h2}
r^d=\left (1-\frac{dGM}{\xi R^d}t\right )a^d.
\end{equation}
Therefore, the particles reach $r=0$ at a time
\begin{equation}
\label{h3}
t_*=\frac{\xi R^d}{dGM},
\end{equation}
whatever their initial position $a$. This leads to a complete collapse of the
system (Dirac peak) at a time $t=t_*$. According to Eqs. (\ref{f9}), (\ref{h1}) and (\ref{h2}), the mass profile at time $t$ is
\begin{equation}
\label{h4}
M(r,t)=\frac{M}{R^d}\frac{r^d}{1-\frac{t}{t_*}}.
\end{equation}
The sphere of particles remains spatially homogeneous with a radius decreasing in time like
\begin{equation}
\label{h5}
R(t)=R\left (1-\frac{t}{t_*}\right )^{1/d}.
\end{equation}
The density increases with time like
\begin{equation}
\label{h6}
\rho(t)=\frac{\rho(0)}{1-\frac{t}{t_*}}.
\end{equation}
This result can be directly obtained from Eqs. (\ref{f1})-(\ref{f2}) using the fact that the density is spatially homogeneous. Indeed, using the Poisson equation (\ref{f2}), Eq. (\ref{f1}) becomes
\begin{equation}
\label{h7}
\frac{d\rho}{dt}=\frac{S_d G}{\xi}\rho^2,
\end{equation}
which leads to Eq. (\ref{h6}) after integration.

{\it Remark:} in the context of the Euler-Poisson
system at $T=0$, the free-fall time of a homogeneous sphere is
$t_{ff}=(3\pi/32 G\rho_0)^{1/2}$ (in $d=3$) \cite{hunter}. This is of
the order of the dynamical time $t_{dyn}=(3\pi/16 G\rho_0)^{1/2}$
\cite{bt}. For the Smoluchowski-Poisson system at $T=0$, the collapse time
(\ref{h3}) of a homogeneous sphere can be written $t_{*}=\xi/S_d
\rho_0 G$. It is of the order $\xi t_{dyn}^2$.

\section{The pre-collapse of a parabolic profile}
\label{sec_pre}

\subsection{The exact solution}
\label{sec_preexact}

We now consider an initial mass profile of the form
\begin{equation}
\label{p1}
M(a,0)=A(a^d-B a^{d+2}),
\end{equation}
where $A$ and $B$ are two positive constants. Using Eq. (\ref{f8}), we find
that the corresponding density profile is parabolic
\begin{equation}
\label{p2}
\rho(a,0)=\frac{dA}{S_d}\left (1-\frac{d+2}{d} B a^2\right ).
\end{equation}
These expressions are valid for $a\le R$ where $R$ is the radius at
which the density vanishes.  The constant $B$ is related to the radius
$R$ of the initial configuration by
\begin{equation}
\label{p3}
B=\frac{d}{(d+2)R^2}.
\end{equation}
On the other hand, the constant $A$ is related to the total mass $M$ of the system, and to its radius $R$,  by
\begin{equation}
\label{p4}
A=\frac{d+2}{2}\frac{M}{R^d}.
\end{equation}
According to Eqs. (\ref{f11})
and (\ref{p1}), the position at time $t$ of a particle initially located at
$r=a$ is
\begin{equation}
\label{p5}
r^d=a^d-\frac{d}{\xi}G A a^d (1-B a^2) t.
\end{equation}
The time at which the particle initially located at $r=a>0$ reaches the origin $r=0$ is
\begin{equation}
\label{p6}
t_*(a)=\frac{\xi}{dGA(1-Ba^2)}.
\end{equation}
In particular, the time at which the last particle, i.e. the one initially located at $a=R$, reaches the origin is
\begin{equation}
\label{p7}
t_{end}=\frac{(d+2)\xi}{2dGA}=\frac{\xi R^d}{dGM}.
\end{equation}
This is the final time since, at that time, all the particles have
collapsed at the origin. The density profile is a Dirac peak
$\rho=M\delta({\bf r})$ containing all the mass and there is no
further evolution. As may appear surprising at first sight, the time
at which the first particle, i.e. the one initially located at
$a=0^+$, reaches the origin is {\it finite}:
\begin{equation}
\label{p8}
t_{coll}=\frac{\xi}{dGA}=\frac{2\xi R^d}{d(d+2)GM}.
\end{equation}
This is the time at which the density profile becomes singular at the
origin, since an infinitesimal number of particles has reached
$r=0$. This corresponds to a finite time singularity in which the
central density diverges. We note that
\begin{equation}
\label{p9}
t_{end}=\frac{d+2}{2}t_{coll}.
\end{equation}
Combining the foregoing equations, the conservation of mass (\ref{f9}) and the equation of motion (\ref{f11}) can be written
\begin{equation}
\label{p11}
M(r,t)=\frac{\xi}{d G t_{coll}}a^d (1-Ba^2),
\end{equation}
\begin{equation}
\label{p12}
r^d=\frac{1}{t_{coll}}(t_{coll}-t)a^d+B\frac{t}{t_{coll}}a^{d+2}.
\end{equation}
These equations, parameterized by $a$, completely determine the
evolution of the mass profile for all times. However, in order to
exhibit the self-similar structure of the solution close to $t_{coll}$
(see Sec. \ref{sec_press}), it is convenient to define
\begin{equation}
\label{p13}
y=\frac{1}{t_{coll}}\frac{a^d}{(t_{coll}-t)^{{d}/{2}}},
\end{equation}
and
\begin{equation}
\label{p14}
C=B t_{coll}^{{(d+2)/d}},
\end{equation}
in which case, Eqs. (\ref{p11}) and (\ref{p12}) can be rewritten
\begin{equation}
\label{p15}
M(r,t)=\frac{\xi}{d G}(t_{coll}-t)^{d/2}y\left (1-C\frac{t_{coll}-t}{t_{coll}}y^{2/d}\right ),
\end{equation}
\begin{equation}
\label{p16}
x^d=y+C\frac{t}{t_{coll}}y^{{(d+2)/}{d}},
\end{equation}
\begin{equation}
\label{p17}
x=\frac{r}{(t_{coll}-t)^{{(d+2)/}{2d}}},
\end{equation}
where $x$ is the appropriate scaling variable and $y$ is just a dummy
variable.  Using Eqs. (\ref{f8}), (\ref{p15}) and (\ref{p16}), the
corresponding density profile is
\begin{eqnarray}
\label{p19}
\rho(r,t)=\frac{\xi}{S_d G}\frac{1}{t_{coll}-t}\left\lbrack 1-\frac{C}{t_{coll}}(t_{coll}-t)\frac{2+d}{d}y^{2/d}\right\rbrack\nonumber\\
\times\frac{1}{\frac{d+2}{d} C\frac{t}{t_{coll}}y^{2/d}+1}.\qquad
\end{eqnarray}
The central density behaves like
\begin{eqnarray}
\label{p20}
\rho(0,t)=\frac{\xi}{S_d G}\frac{1}{t_{coll}-t},
\end{eqnarray}
and it diverges at $t=t_{coll}$. Finally, according to Eq. (\ref{p12}), the position at time $t$ of the last particle, i.e. the one located at $a=R$ at time $t=0$, is
\begin{equation}
\label{p18}
r_{max}(t)^d=\frac{1}{t_{coll}}(t_{coll}-t)R^d+B\frac{t}{t_{coll}}R^{d+2}.
\end{equation}
We can check that $\rho(r_{max},t)=0$. Let us now consider particular limits of these equations.

For $t\rightarrow 0$, we obtain after some calculations
\begin{eqnarray}
\label{p21}
\rho(r,t)\simeq \rho(r,0)+\frac{\xi}{S_d G t_{coll}}\frac{t}{t_{coll}}\biggl\lbrack 1-B\frac{2(d+1)(d+2)}{d^2}r^2\nonumber\\
+B^2\frac{(d+2)(d+4)}{d^2}r^4\biggr\rbrack.\qquad
\end{eqnarray}
The corresponding mass profile is
\begin{eqnarray}
\label{p21b}
M(r,t)\simeq M(r,0)+\frac{\xi}{G t_{coll}}\frac{t}{t_{coll}}\biggl\lbrack \frac{r^d}{d}-B\frac{2(d+1)}{d^2}r^{d+2}\nonumber\\
+B^2\frac{d+2}{d^2}r^{d+4}\biggr\rbrack.\qquad
\end{eqnarray}
This is valid for $r\le r_{max}(t)$ with
\begin{eqnarray}
\label{p22}
r_{max}(t)\simeq R\left\lbrack 1-\frac{2}{d(d+2)}\frac{t}{t_{coll}}\right\rbrack.
\end{eqnarray}
For $t\rightarrow 0$, the central density behaves like
\begin{eqnarray}
\label{p23}
\rho(0,t)\simeq \frac{\xi}{S_d G t_{coll}}\left (1+\frac{t}{t_{coll}}\right ).
\end{eqnarray}

Taking $t=t_{coll}$ in Eqs. (\ref{p15})-(\ref{p17}), and using $x^d=Cy^{(d+2)/d}$, we obtain the exact profiles
\begin{eqnarray}
\label{px1}
M(r,t_{coll})=\frac{\xi}{d G}\frac{1}{C^{{d/}{(d+2)}}}r^{{d^2/}{(d+2)}}\nonumber\\
\times\left \lbrack 1-\frac{C^{{d/}{(d+2)}}}{t_{coll}}r^{{2d/}{(d+2)}}\right \rbrack,
\end{eqnarray}
\begin{eqnarray}
\label{px2}
\rho(r,t_{coll})=\frac{\xi}{S_d G}\frac{d}{d+2}\frac{1}{C^{{d/}{(d+2)}}}\frac{1}{r^{{2d/}{(d+2)}}}\nonumber\\
\times\left \lbrack 1-\frac{d+2}{d}\frac{C^{{d/}{(d+2)}}}{t_{coll}}r^{{2d/}{(d+2)}}\right \rbrack.
\end{eqnarray}
According to Eqs. (\ref{p18}) and (\ref{p14}), the position at time $t_{coll}$ of the last particle is
\begin{equation}
\label{px3}
r_{max}=\frac{C^{1/d}}{t_{coll}^{{(d+2)/}{d^2}}}R^{{(d+2)/}{d}}.
\end{equation}
We can check that $M(r_{max},t_{coll})=M$. More precisely, using Eqs. (\ref{px1}) and  (\ref{px2}), we find that
\begin{equation}
\label{px4}
M(r,t_{coll})\simeq M\left\lbrack 1-\frac{d^3}{2(d+2)}\left (\frac{r_{max}-r}{r_{max}}\right )^2\right \rbrack,
\end{equation}
\begin{equation}
\label{px4b}
\rho(r,t_{coll})\simeq \frac{d^2 M}{S_d R^d}\left (1-\frac{r}{r_{max}}\right ),
\end{equation}
for $r\rightarrow r_{max}$. The behavior of $M(r,t_{coll})$ and $\rho(r,t_{coll})$ for $r\rightarrow 0$ is given in the following section.

\subsection{Self-similar solution for $t\rightarrow t_{coll}$ and $r\rightarrow 0$}
\label{sec_press}

For $t\rightarrow t_{coll}$ and $r\rightarrow 0$, Eqs. (\ref{p15})-(\ref{p17}) reduce to the form
\begin{equation}
\label{ps1}
M(r,t)=\frac{\xi}{d G}(t_{coll}-t)^{d/2}y,
\end{equation}
\begin{equation}
\label{ps2}
x^d=y+Cy^{{(d+2)/}{d}},
\end{equation}
\begin{equation}
\label{ps3}
x=\frac{r}{(t_{coll}-t)^{{(d+2)/}{2d}}},
\end{equation}
showing that the evolution becomes self-similar as we approach the collapse time. The corresponding  density profile is
\begin{equation}
\label{ps4}
\rho(r,t)=\frac{\xi}{S_d G}\frac{1}{t_{coll}-t}\frac{1}{1+\frac{d+2}{d}C y^{2/d}}.
\end{equation}
The central density increases like
\begin{equation}
\label{ps4b}
\rho(0,t)=\frac{\xi}{S_d G}\frac{1}{t_{coll}-t},
\end{equation}
and diverges at the collapse time. In parallel, the typical core radius
\begin{eqnarray}
\label{p24}
r_0(t)\sim (t_{coll}-t)^{(d+2)/2d}.
\end{eqnarray}
appearing in Eq. (\ref{ps3}) decreases and tends to zero at $t=t_{coll}$.
Comparing Eq.  (\ref{p24}) with Eq. (\ref{ps4b}), we find that the core radius is related to the central density $\rho_0(t)=\rho(0,t)$ by the relation
\begin{eqnarray}
\label{p25}
\rho_0 r_0^{\alpha}\sim 1,
\end{eqnarray}
involving the scaling exponent
\begin{eqnarray}
\label{p26}
\alpha=\frac{2d}{d+2}.
\end{eqnarray}
Note also that the mass within the core radius is
\begin{eqnarray}
\label{p27}
M(r_0(t),t)\sim \rho_0(t)r_0^d(t)\sim (t_{coll}-t)^{d/2},
\end{eqnarray}
and it tends to zero as $t\rightarrow t_{coll}$. Therefore, there is no Dirac peak in the pre-collapse regime.

For $t=t_{coll}$ and $r\rightarrow 0$, Eq. (\ref{ps2}) reduce to $x^d=Cy^{{(d+2)/}{d}}$ and we obtain the power-law profiles
\begin{equation}
\label{ps5}
M(r,t_{coll})=\frac{\xi}{d G}\frac{1}{C^{{d/}{(d+2)}}}r^{{d^2/}{(d+2)}},
\end{equation}
\begin{equation}
\label{ps6}
\rho(r,t_{coll})=\frac{\xi}{S_d G}\frac{d}{d+2}\frac{1}{C^{{d/}{(d+2)}}}\frac{1}{r^{{2d/}{(d+2)}}}.
\end{equation}
They can also be derived from Eqs. (\ref{px1}) and (\ref{px2}) for $r\rightarrow 0$. This shows that the system develops a finite time singularity: the density profile diverges at the origin and scales like $r^{-{2d/}{(d+2)}}$. By contrast, the mass $M(r,t_{coll})$ tends to zero at the origin. This means that only an {\it infinitesimal} number of particles has reached the origin at $t=t_{coll}$, leading to infinite central density but zero central  mass.

\subsection{The half central density radius}

Using Eq. (\ref{p19}), the radius $r_*(t)$ corresponding to half the central density, i.e. such that $\rho(r_*(t),t)=\rho(0,t)/2$, is
\begin{eqnarray}
r_*(t)=\left (\frac{d}{d+2}\frac{1}{C}\right )^{1/2}\frac{(t_{coll}-t)^{(d+2)/2d}}{\left (2-\frac{t}{t_{coll}}\right )^{(d+2)/2d}}\nonumber\\
\times\left (2-\frac{2}{d+2}\frac{t}{t_{coll}}\right )^{1/d}.
\label{fk}
\end{eqnarray}
For $t=0$, we obtain
\begin{eqnarray}
\label{a7}
r_{*}(0)=\frac{R}{\sqrt{2}},
\end{eqnarray}
a result that can be directly obtained from Eq. (\ref{p2}). For $t\rightarrow 0$, we find that
\begin{eqnarray}
\label{a7b}
r_{*}(t)\simeq \frac{R}{\sqrt{2}}\left \lbrack 1-\frac{d^2+4d+8}{4d(d+2)}\frac{t}{t_{coll}}\right \rbrack.
\end{eqnarray}
For $t=t_{coll}$, Eq. (\ref{fk})  gives $r_*=0$. In the self-similar regime $t\rightarrow t_{coll}$, we find that
\begin{eqnarray}
r_*(t)={\left (\frac{d}{d+2}\frac{1}{C}\right )^{1/2}}\left\lbrack\frac{2(d+1)}{d+2}\right \rbrack^{1/d} (t_{coll}-t)^{(d+2)/2d}.\nonumber\\
\end{eqnarray}
In the self-similar regime, the half central density radius $r_*(t)$
behaves like the typical core radius $r_0(t)$, see
Eq. (\ref{p24}). The mass within the sphere of radius $r_*$ evolves
with time like
\begin{eqnarray}
\label{a6nr}
M(r_*(t),t)\sim \rho_0(t)r_*(t)^d\nonumber\\
\sim \frac{2-\frac{2}{d+2}\frac{t}{t_{coll}}}{\left (2-\frac{t}{t_{coll}}\right )^{{(d+2)/}{2}}}(t_{coll}-t)^{d/2}.
\end{eqnarray}
In the self-similar regime, we find that $M(r_*(t),t)$ tends to zero like $\sim (t_{coll}-t)^{d/2} \sim M(r_0(t),t)$.

\subsection{More explicit solutions for $d=2$}
\label{sec_deux}

In $d=2$, Eq. (\ref{p16}) can be solved explicitly to obtain $y(x,t)$. Therefore, the exact solution can be written
\begin{equation}
\label{d1}
M(r,t)=\frac{\xi}{2G}(t_{coll}-t) y\left (1-C\frac{t_{coll}-t}{t_{coll}}y\right ),
\end{equation}
\begin{eqnarray}
\label{d1b}
\rho(r,t)=\frac{\xi}{2\pi G}\frac{1}{t_{coll}-t}\left\lbrack 1-\frac{2C}{t_{coll}}(t_{coll}-t)y\right\rbrack
\frac{1}{\frac{2Ct}{t_{coll}}y+1},\nonumber\\
\end{eqnarray}
\begin{equation}
\label{d2}
y=\frac{-1+\sqrt{1+4C\frac{t}{t_{coll}}x^2}}{\frac{2Ct}{t_{coll}}},
\end{equation}
\begin{equation}
\label{d3}
x=\frac{r}{t_{coll}-t}.
\end{equation}
For $t\rightarrow t_{coll}$ and $r\rightarrow 0$, the self-similar solution can be written
\begin{equation}
\label{d4}
M(r,t)=\frac{\xi}{2G}(t_{coll}-t) y,
\end{equation}
\begin{eqnarray}
\label{d'B}
\rho(r,t)=\frac{\xi}{2\pi G}\frac{1}{t_{coll}-t}
\frac{1}{2C y+1},
\end{eqnarray}
\begin{equation}
\label{d5}
y=\frac{-1+\sqrt{1+4Cx^2}}{2C},
\end{equation}
\begin{equation}
\label{d6}
x=\frac{r}{t_{coll}-t}.
\end{equation}

\section{The post-collapse regime}
\label{sec_post}

The previous solution shows that the system forms a finite time singularity: the central density becomes infinite in a finite time $t=t_{coll}$. However, the mass contained within a sphere of radius $\epsilon$ tends to zero as $\epsilon\rightarrow 0$. This is due to the fact that only an {\it infinitesimal} number  of particles has reached $r=0$ at $t=t_{coll}$ leading to infinite central density but zero central mass. This cannot be the final equilibrium state of the system since statistical mechanics predicts that the equilibrium state of a self-gravitating gas in the canonical ensemble is a Dirac peak containing all the particles (this is the configuration that makes the free energy tend to $-\infty$ due to the divergence of energy) \cite{kiessling,aaiso}. This is true at any temperature. For a self-gravitating Brownian gas a $T=0$, we have previously established that the Dirac peak $\rho({\bf r})=M\delta({\bf r})$ forms at a time $t_{end}$ given by Eq. (\ref{p7}). Therefore, the evolution continues in the post collapse regime $t_{coll}\le t\le t_{end}$ where a Dirac peak grows and accretes progressively all the surrounding particles. The exact description of this process for $T=0$ is the object of the present section.

\subsection{The exact solution}
\label{sec_postexact}

For $t\ge t_{coll}$, the equation of motion (\ref{f11}) can be written
\begin{equation}
\label{po1}
r^d=a^d-\frac{d}{\xi}GM(a,t_{coll})(t-t_{coll}),
\end{equation}
where $M(a,t_{coll})$ is the mass profile at time $t=t_{coll}$. Using Eq. (\ref{px1}), we obtain
\begin{eqnarray}
\label{po2}
r^d=a^d-\frac{1}{C^{{d/}{(d+2)}}}a^{{d^2/}{(d+2)}}\nonumber\\
\times\left \lbrack 1-\frac{C^{{d/}{(d+2)}}}{t_{coll}}a^{{2d/}{(d+2)}}\right \rbrack (t-t_{coll}).
\end{eqnarray}
At $t=t_{coll}+\Delta t$, the mass contained within the sphere of radius $a_*$ at $t=t_{coll}$ has reached $r=0$. From Eq. (\ref{po2}), we get
\begin{equation}
\label{po3}
a_*^{{2d/}{(d+2)}}=\frac{\Delta t}{C^{{d/}{(d+2)}}}\frac{1}{1+\frac{\Delta t}{t_{coll}}}.
\end{equation}
This leads to the formation of a Dirac peak of mass $M_{D}(t)=M(a_*,t_{coll})$. Substituting Eq. (\ref{po3}) in Eq. (\ref{px1}), we obtain
\begin{equation}
\label{po4}
M_{D}(t)=\frac{\xi}{dG}\frac{1}{C^{d/2}}\left (\frac{t_{coll}}{t}\right )^{{(d+2)/}{2}}(t-t_{coll})^{d/2}.
\end{equation}
We can check that $M_{D}(t_{end})=M$ so that all the mass has been absorbed in the Dirac peak at $t=t_{end}$. More precisely, we find that
\begin{equation}
\label{po4b}
M_{D}(t)\simeq M\left\lbrack 1-\frac{d+2}{2d}\left (1-\frac{t}{t_{end}}\right )^2\right\rbrack,
\end{equation}
for $t\rightarrow t_{end}$. It is relevant to write the conservation of mass (\ref{f9}) in the form
\begin{equation}
\label{po5}
M_{tot}(r,t)\equiv M_{D}(t)+M(r,t)=M(a,t_{coll}),
\end{equation}
where $M_{D}(t)$ is the mass contained in the Dirac peak and $M(r,t)$ is the mass exterior to the Dirac. Similarly, we write the density profile as
\begin{equation}
\label{po6}
\rho_{tot}({\bf r},t)= M_{D}(t)\delta({\bf r})+\rho({\bf r},t).
\end{equation}
The regular part of the profile $\rho({\bf r},t)$ is  the residual density defined as the density after the central peak has been subtracted. Using Eq. (\ref{px1}), the evolution of the mass profile in the post-collapse regime is given by
\begin{eqnarray}
\label{po7}
M_{D}(t)+M(r,t)=\frac{\xi}{d G}\frac{1}{C^{{d/}{(d+2)}}}a^{{d^2/}{(d+2)}}\nonumber\\
\times\left \lbrack 1-\frac{C^{{d/}{(d+2)}}}{t_{coll}}a^{{2d/}{(d+2)}}\right \rbrack,
\end{eqnarray}
\begin{eqnarray}
\label{po8}
r^d=a^d-\frac{1}{C^{{d/}{(d+2)}}}a^{{d^2/}{(d+2)}}\nonumber\\
\times\left \lbrack 1-\frac{C^{{d/}{(d+2)}}}{t_{coll}}a^{{2d/}{(d+2)}}\right \rbrack (t-t_{coll}).
\end{eqnarray}
These equations completely determine the evolution of the mass profile in the post-collapse regime. In order to exhibit the self-similar structure of the solution close to $t_{coll}$ (see Sec. \ref{sec_postss}), it is convenient to define
\begin{equation}
\label{po9}
y=\frac{1}{C^{{d/}{(d+2)}}}\frac{a^{{d^2/}{(d+2)}}}{(t-t_{coll})^{{d/}{2}}},
\end{equation}
in which case, Eqs. (\ref{po7}) and (\ref{po8}) can be rewritten
\begin{eqnarray}
\label{po10}
M_{D}(t)+M(r,t)=\frac{\xi}{d G}(t-t_{coll})^{d/2}\nonumber\\
\times y\left (1-C\frac{t-t_{coll}}{t_{coll}}y^{2/d}\right ),
\end{eqnarray}
\begin{equation}
\label{po11}
x^d=C\frac{t}{t_{coll}}y^{{(d+2)/}{d}}-y,
\end{equation}
\begin{equation}
\label{po12}
x=\frac{r}{(t-t_{coll})^{{(d+2)/}{2d}}}.
\end{equation}
Note the similarities and the differences with the pre-collapse solution (\ref{p15})-(\ref{p17}). If we subtract the contribution of the Dirac peak, using Eq. (\ref{po4}), we obtain
\begin{eqnarray}
\label{po13}
M(r,t)=\frac{\xi}{d G}(t-t_{coll})^{d/2}\nonumber\\
\times\biggl\lbrack y\left (1-C\frac{t-t_{coll}}{t_{coll}}y^{2/d}\right )
-\frac{1}{C^{d/2}}\left (\frac{t_{coll}}{t}\right )^{{(d+2)/}{2}}\biggr\rbrack,\nonumber\\
\end{eqnarray}
\begin{equation}
\label{po14}
x^d=C\frac{t}{t_{coll}}y^{{(d+2)/}{d}}-y,
\end{equation}
\begin{equation}
\label{po15}
x=\frac{r}{(t-t_{coll})^{{(d+2)/}{2d}}}.
\end{equation}
Using Eqs. (\ref{f8}), (\ref{po13}) and (\ref{po14}), the corresponding density profile is
\begin{eqnarray}
\label{a1}
\rho(r,t)=\frac{\xi}{S_d G}\frac{1}{t-t_{coll}}\left\lbrack 1-\frac{C}{t_{coll}}(t-t_{coll})\frac{2+d}{d}y^{2/d}\right\rbrack\nonumber\\
\times\frac{1}{\frac{d+2}{d} C\frac{t}{t_{coll}}y^{2/d}-1}.\qquad
\end{eqnarray}
According to Eqs. (\ref{a1}), (\ref{po14}) and (\ref{po15}), the central residual density decreases with time like
\begin{eqnarray}
\label{a4}
\rho(0,t)=\frac{\xi}{S_d G}\frac{1}{t-t_{coll}}\left (\frac{t_{end}}{t}-1\right ).
\end{eqnarray}
Let us now consider particular limits of these equations.

For $t=t_{coll}$, we recover the results (\ref{px1})-(\ref{px4b}). For $t=t_{end}$, using Eq. (\ref{p18}), we can check  that $r_{max}(t_{end})=0$.
 More precisely, for $t\rightarrow t_{end}$, we obtain after some calculations
\begin{eqnarray}
\label{a2}
\rho(r,t)\simeq \frac{2\xi}{dS_d G}\frac{1}{t_{coll}}\biggl \lbrack \frac{t_{end}-t}{t_{end}}-C^{d/2}\left (\frac{d+2}{d}\right )^{d/2}\frac{r^{d}}{t_{coll}^{{(d+2)/}{2}}}\biggr \rbrack.\nonumber\\
\end{eqnarray}
The corresponding mass profile is
\begin{eqnarray}
\label{a2b}
M(r,t)\simeq \frac{2\xi r^d}{d^2 G t_{coll}}\biggl \lbrack \frac{t_{end}-t}{t_{end}}-\left (\frac{d+2}{d}\right )^{d/2}\frac{r^{d}}{2}\frac{C^{d/2}}{t_{coll}^{{(d+2)/}{2}}}\biggr \rbrack.\nonumber\\
\end{eqnarray}
This is valid for $r\le r_{max}(t)$ with
\begin{eqnarray}
\label{a3}
r_{max}\simeq \left (\frac{t_{end}-t}{t_{end}}\right )^{1/d}R.
\end{eqnarray}
The total mass outside the Dirac is $M(t)=M(r_{max},t)$. Using Eqs. (\ref{a2b}) and (\ref{a3}), we obtain
\begin{eqnarray}
\label{a2c}
M(t)\simeq M\frac{d+2}{2d} \left (\frac{t_{end}-t}{t_{end}}\right )^2.
\end{eqnarray}
Comparing with Eq. (\ref{po4b}), we check that $M_D(t)+M(t)=M$, as it should. On the other hand, for $t\rightarrow t_{end}$, the central density (\ref{a4}) decreases like
\begin{eqnarray}
\label{a5}
\rho(0,t)\simeq \frac{2\xi}{dS_d G}\frac{1}{t_{coll}}\left (1-\frac{t}{t_{end}}\right ),
\end{eqnarray}
in agreement with Eq. (\ref{a2}).

\subsection{Self-similar solution for $t\rightarrow t_{coll}$ and $r\rightarrow 0$}
\label{sec_postss}

As the Dirac peak grows in the post-collapse regime, the central residual density decreases. Soon after $t=t_{coll}$, the residual density profile $\rho(r,t)$ follows a  post-collapse self-similar evolution, reverse to the pre-collapse self-similar evolution, during which the central residual density $\rho(0,t)$ decreases while the core radius $r_0(t)$ increases. This is the object of the present subsection.

For $t\rightarrow t_{coll}$, the mass of the Dirac grows like
\begin{equation}
\label{pos3b}
M_{D}(t)=\frac{\xi}{dG}\frac{1}{C^{d/2}}(t-t_{coll})^{d/2}.
\end{equation}
Furthermore, for $t\rightarrow t_{coll}$ and $r\rightarrow 0$, Eqs. (\ref{po13})-(\ref{po15}) reduce to the form
\begin{eqnarray}
\label{pos1}
M(r,t)=\frac{\xi}{d G}(t-t_{coll})^{d/2}\left (y-\frac{1}{C^{d/2}}\right ),
\end{eqnarray}
\begin{equation}
\label{pos2}
x^d=Cy^{{(d+2)/}{d}}-y,
\end{equation}
\begin{equation}
\label{pos3}
x=\frac{r}{(t-t_{coll})^{{(d+2)/}{2d}}},
\end{equation}
showing that the evolution is self-similar just after the collapse time.
The corresponding  density profile is
\begin{eqnarray}
\label{pos4}
\rho(r,t)=\frac{\xi}{S_d G}\frac{1}{t-t_{coll}}\frac{1}{\frac{d+2}{d}Cy^{2/d}-1}.
\end{eqnarray}
The central residual density behaves like
\begin{eqnarray}
\label{pos5}
\rho(0,t)=\frac{d\xi}{2S_d G}\frac{1}{t-t_{coll}}.
\end{eqnarray}
Therefore, the central residual density decreases as more and more mass is absorbed in the Dirac peak.
In parallel, the typical core radius
\begin{eqnarray}
\label{pos6}
r_0(t)\sim (t-t_{coll})^{(d+2)/2d},
\end{eqnarray}
appearing in Eq. (\ref{pos3}) increases.
Comparing Eq. (\ref{pos6}) with Eq. (\ref{pos5}), we find that the core radius is related to the central density $\rho_0(t)=\rho(0,t)$ by the relation
\begin{eqnarray}
\label{pos7}
\rho_0 r_0^{\alpha}\sim 1,
\end{eqnarray}
involving the scaling exponent
\begin{eqnarray}
\label{pos8}
\alpha=\frac{2d}{d+2},
\end{eqnarray}
the same as in the pre-collapse regime. The mass within the core radius is
\begin{eqnarray}
\label{pos9}
M(r_0(t),t)\sim \rho_0(t)r_0^d(t)\sim (t-t_{coll})^{d/2},
\end{eqnarray}
and it increases with time.  Note that the mass (\ref{pos3b}) contained in the Dirac peak has a similar scaling.

\subsection{The half central density radius}
\label{sec_hcdr}

In the self-similar regime, the core radius $r_0(t)$ increases with time indicating that the residual density profile expands as the central density decreases. However, since the system size $r_{max}(t)$ decreases to zero for $t=t_{end}$, the expansion of the density profile cannot be valid for all times. In fact, the core radius $r_0(t)$ makes sense only during the self-similar regime. To study the evolution of the residual density profile in a more general setting, we introduce  the half central density radius $r_*(t)$.

Using Eq. (\ref{a1}), the radius $r_*(t)$ corresponding to half the central density, i.e. such that $\rho(r_*(t),t)=\rho(0,t)/2$, is
\begin{eqnarray}
\label{a6}
r_*(t)=\left (\frac{t_{coll}}{C}\frac{d}{d+2}\right )^{1/2}\frac{\left (\frac{1}{2}+\frac{2+d}{4}\frac{t_{coll}}{t}\right )^{1/2}}{\left (\frac{d-2}{4}t_{coll}+\frac{t}{2}\right )^{{(d+2)/}{2d}}}\nonumber\\
\times\left (\frac{1}{2}t_{coll}-\frac{t}{d+2}\right )^{1/d}(t-t_{coll})^{{(d+2)/}{2d}}.
\end{eqnarray}
For $t=t_{coll}$, this formula gives $r_*=0$. In the self-similar regime $t\rightarrow t_{coll}$, we have
\begin{eqnarray}
\label{a6b}
r_*(t)=\frac{\left (\frac{4+d}{d+2}\right )^{1/2}\left (\frac{2}{d+2}\right )^{1/d}}{C^{1/2}}   (t-t_{coll})^{{(d+2)/}{2d}}.\nonumber\\
\end{eqnarray}
For $t\rightarrow t_{end}$, we obtain
\begin{eqnarray}
\label{a7v}
r_{*}^d(t)\simeq \frac{1}{C^{d/2}}\left (\frac{d}{d+2}\right )^{d/2}\frac{t_{end}-t}{2t_{end}}t_{coll}^{{(d+2)/}{2}},
\end{eqnarray}
a result that can also be directly obtained  from Eq. (\ref{a2}). In the self-similar regime, we find that $r_*(t)\propto (t-t_{coll})^{{(d+2)/}{2d}}$ behaves like $r_0(t)$ so that the density profile spreads. Then, $r_*(t)$ decreases and behaves like $r_{*}\propto (t_{end}-t)^{1/d}$ for $t\rightarrow t_{end}$ so that the density profile shrinks. The mass within the sphere of radius $r_*$ evolves with time like
\begin{eqnarray}
\label{a6c}
M(r_*(t),t)\sim \rho_0(t)r_*(t)^d\qquad\qquad\nonumber\\
\sim \frac{(t_{end}-t)^2}{t}\frac{\left (\frac{1}{2}+\frac{2+d}{4}\frac{t_{coll}}{t}\right )^{d/2}}{\left (\frac{d-2}{4}t_{coll}+\frac{t}{2}\right )^{{(d+2)/}{2}}}(t-t_{coll})^{d/2}.
\end{eqnarray}
In the self-similar regime, we find that $M(r_*(t),t)$ increases like $\sim (t-t_{coll})^{d/2} \sim M(r_0(t),t)$. Then, $M(r_*(t),t)$ decreases and behaves like $M(r_*(t),t)\sim  (t_{end}-t)^{2}$ for $t\rightarrow t_{end}$.

\subsection{More explicit solutions for $d=2$}
\label{sec_da}

In $d=2$, Eq. (\ref{po14}) can be solved explicitly to obtain $y(x,t)$. Therefore, the exact solution can be written
\begin{eqnarray}
\label{da1}
M(r,t)=\frac{\xi}{2G}(t-t_{coll})\nonumber\\
\times\biggl\lbrack y\left (1-C\frac{t-t_{coll}}{t_{coll}}y\right )
-\frac{1}{C}\left (\frac{t_{coll}}{t}\right )^{2}\biggr\rbrack,
\end{eqnarray}
\begin{eqnarray}
\label{da2}
\rho(r,t)=\frac{\xi}{2\pi G}\frac{1}{t-t_{coll}}\left\lbrack 1-\frac{2C}{t_{coll}}(t-t_{coll})y\right\rbrack\nonumber\\
\times\frac{1}{2C\frac{t}{t_{coll}}y-1},\qquad
\end{eqnarray}
\begin{equation}
\label{da3}
y=\frac{1+\sqrt{1+4C\frac{t}{t_{coll}}x^2}}{\frac{2Ct}{t_{coll}}},
\end{equation}
\begin{equation}
\label{da4}
x=\frac{r}{t-t_{coll}}.
\end{equation}
For $t\rightarrow t_{coll}$ and $r\rightarrow 0$, the self-similar solution can be written
\begin{eqnarray}
\label{da5}
M(r,t)=\frac{\xi}{2G}(t-t_{coll})\left (y-\frac{1}{C}\right ),
\end{eqnarray}
\begin{eqnarray}
\label{da6}
\rho(r,t)=\frac{\xi}{2\pi G}\frac{1}{t-t_{coll}}\frac{1}{2Cy-1},\qquad
\end{eqnarray}
\begin{equation}
\label{da7}
y=\frac{1+\sqrt{1+4Cx^2}}{{2C}},
\end{equation}
\begin{equation}
\label{da8}
x=\frac{r}{t-t_{coll}}.
\end{equation}

\section{Illustration of the results}
\label{sec_illustration}

We shall now illustrate the previous analytical results by plotting some relevant quantities. For simplicity, we consider a two-dimensional system ($d=2$) \footnote{Note that the dimension $d=2$ is particularly relevant in chemotaxis \cite{ks} and for the dynamics of colloids at fluid interface \cite{colloids}.}   and choose a system of units such that $M=R=t_{coll}=1$. With these conventions, we have $A=2$, $B=1/2$ and $C=1/2$. On the other hand, $\xi/G=4$ and $S_2=2\pi$.

\begin{figure}[!h]
\begin{center}
\includegraphics[clip,scale=0.3]{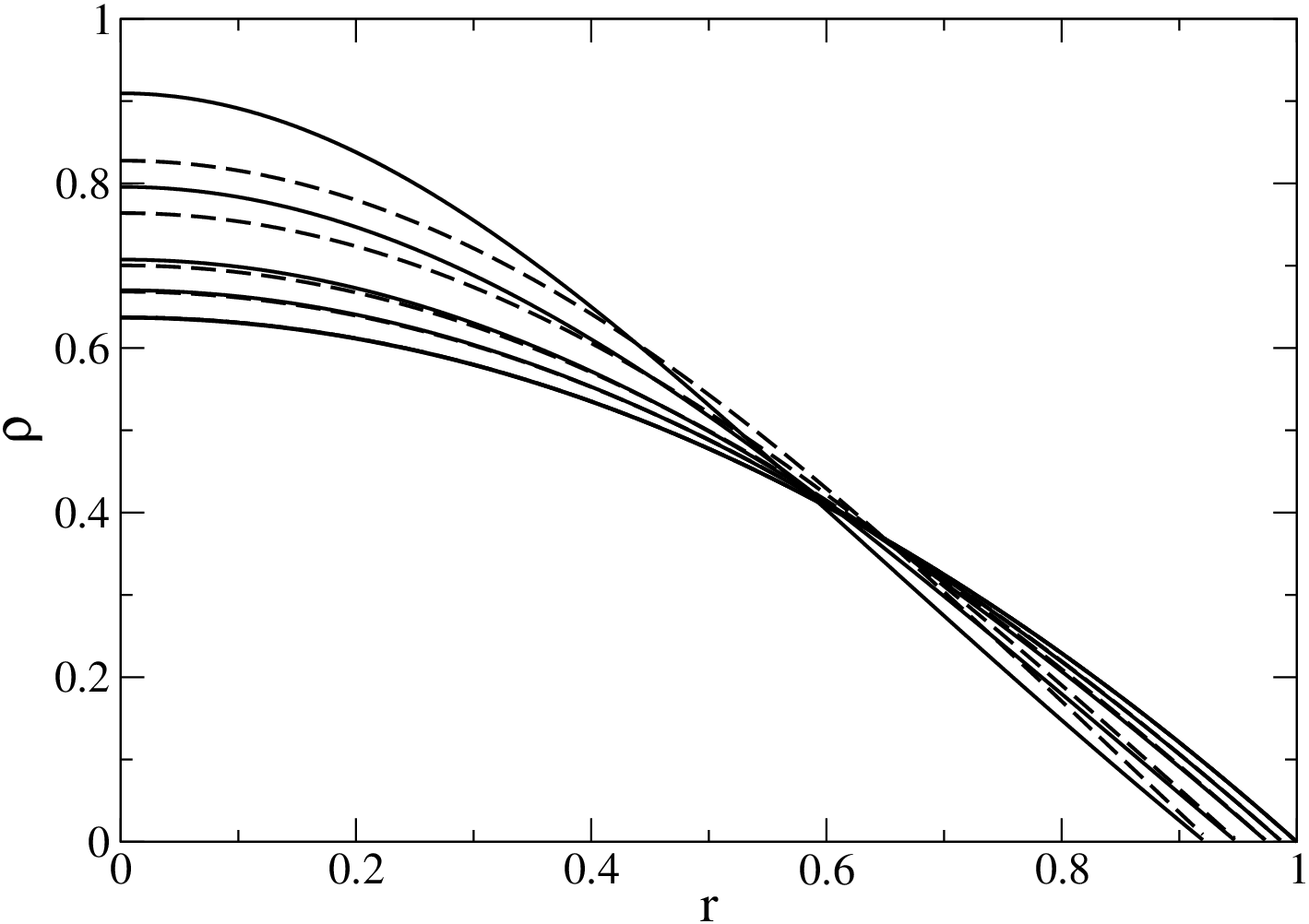}
\caption{Evolution of the density profile $\rho(r,t)$ for small times. We have taken $t=0, 0.05, 0.1, 0.2, 0.3$. The full lines correspond to the exact profile (\ref{u3}) and the dashed lines to the approximation (\ref{u5}) valid for $t\ll 1$.}
\label{densiteTsmall}
\end{center}
\end{figure}

\begin{figure}[!h]
\begin{center}
\includegraphics[clip,scale=0.3]{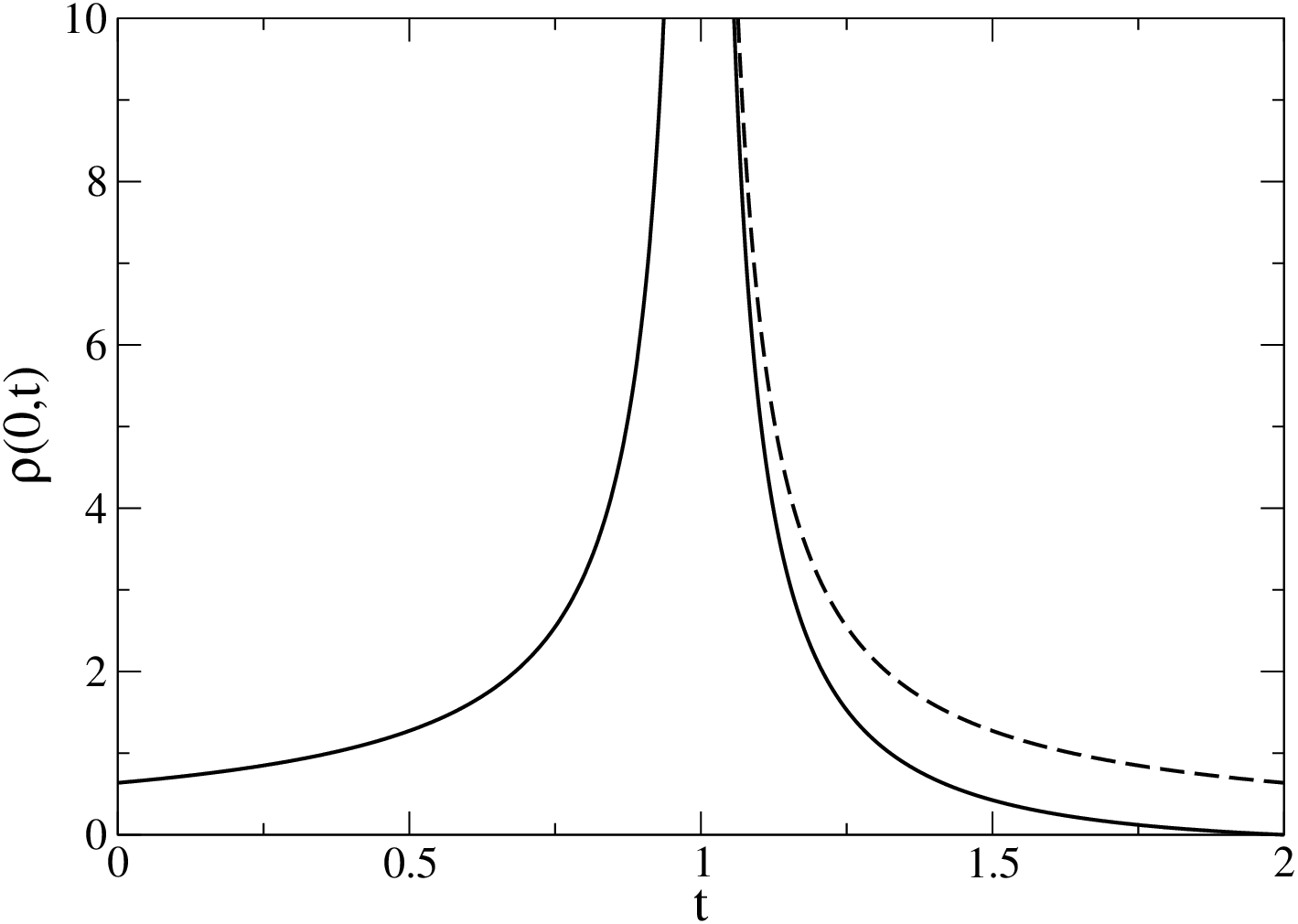}
\caption{Evolution of the central density in the pre and post collapse regimes. The dashed line corresponds to the self-similar solution valid for $t\rightarrow t_{coll}=1$.}
\label{rho0}
\end{center}
\end{figure}

\begin{figure}[!h]
\begin{center}
\includegraphics[clip,scale=0.3]{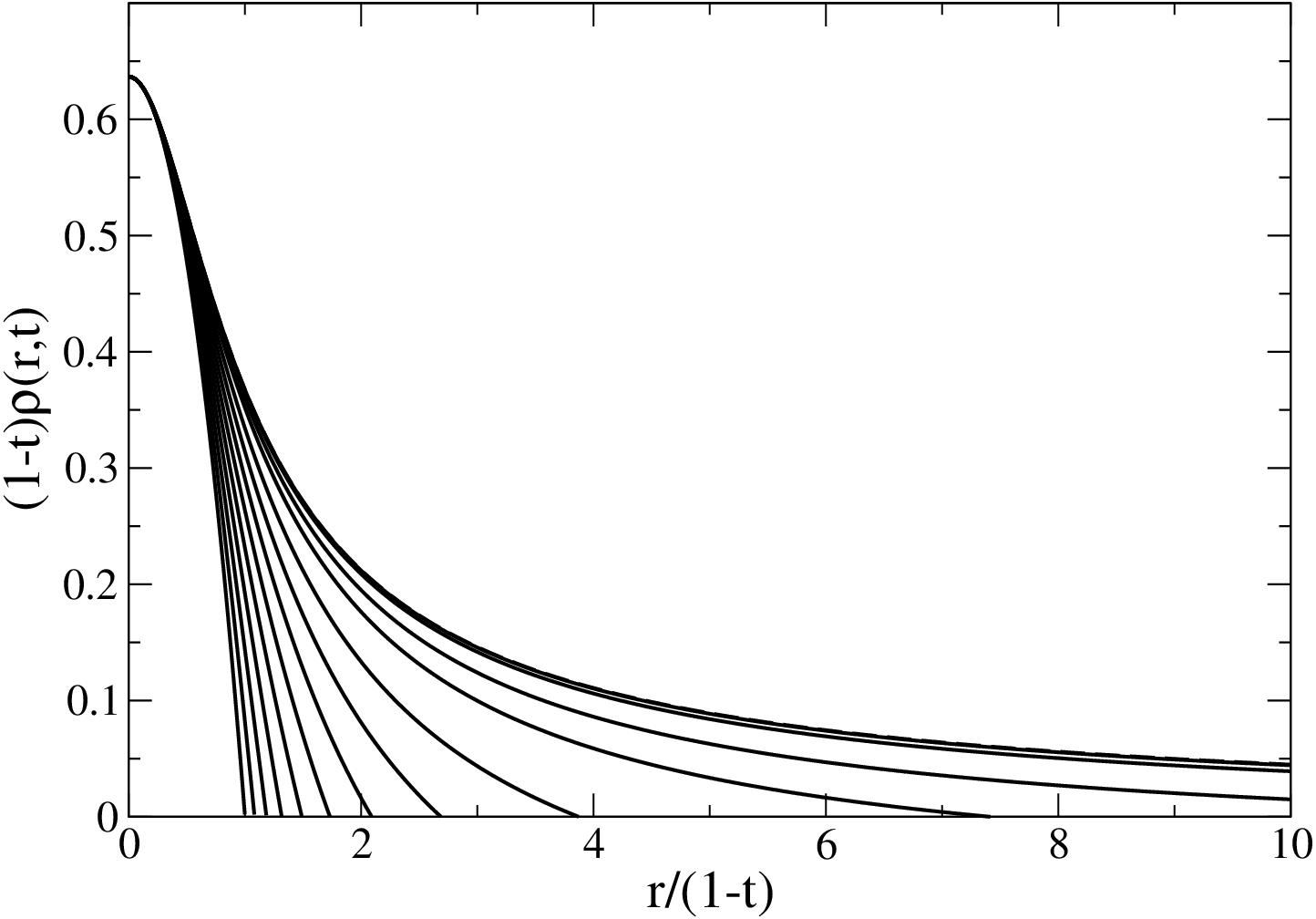}
\caption{Evolution of the density profile $\rho(r,t)$ using self-similar variables. We have taken $t=0, 0.1, 0.2, 0.3, 0.4, 0.5, 0.6, 0.7, 0.8, 0.9, 0.95, 0.99, 0.999$. The full lines correspond to the exact profile (\ref{u3})-(\ref{u4}).  They tend to the invariant profile (\ref{u8}) represented by a dashed line (hardly visible).}
\label{precollapseSS}
\end{center}
\end{figure}

\begin{figure}[!h]
\begin{center}
\includegraphics[clip,scale=0.3]{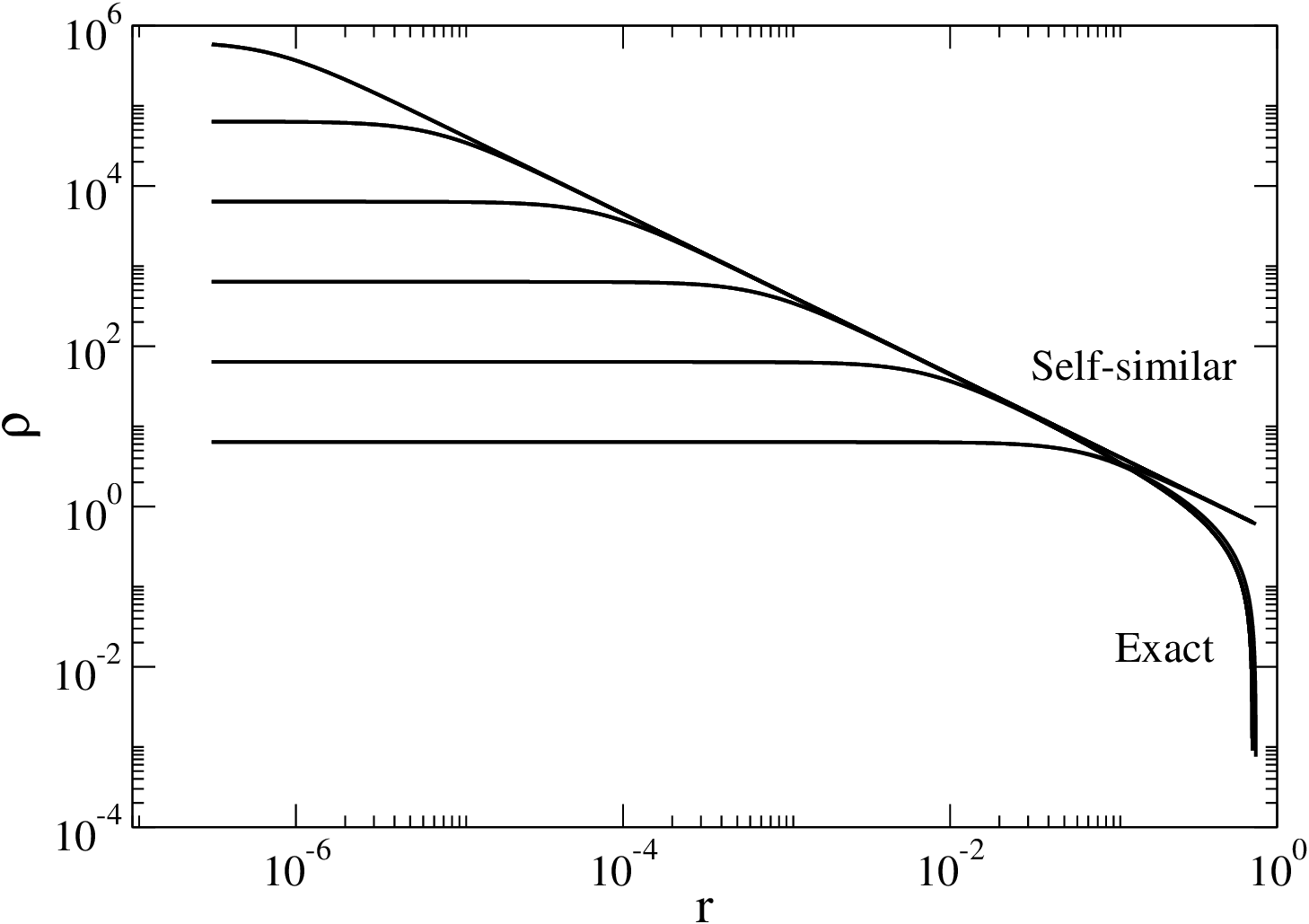}
\caption{Evolution of the density profile $\rho(r,t)$ in log-log plot. We have taken $t_{coll}-t=10^{-6}, 10^{-5}, 10^{-4}, 10^{-3}, 10^{-2}, 10^{-1}$. We have plotted the exact profile (\ref{u3})-(\ref{u4}) and the self-similar profile (\ref{u7})-(\ref{u8}) (dashed lines). At $t=t_{coll}$, the exact profiles tend to the density (\ref{u10}) and the self-similar profiles to the density (\ref{u9}). For sufficiently small $t_{coll}-t$, the self-similar profiles and the exact profiles are in good agreement except for $r\sim r_{max}(t)$. }
\label{densiteLL}
\end{center}
\end{figure}

The initial density profile ($t=0$) is the parabole
\begin{equation}
\label{u1}
\rho(r,0)=\frac{2}{\pi}(1-r^2).
\end{equation}
The finite time singularity occurs at $t=t_{coll}=1$  and the Dirac peak containing the whole mass is formed at $t=t_{end}=2$. The size of the system, i.e. the radius at which the density vanishes, decreases like
\begin{equation}
\label{u2}
r_{max}(t)=\left (1-\frac{t}{2}\right )^{1/2}.
\end{equation}
The exact density profile for $0\le t\le t_{coll}=1$ (pre-collapse) is
\begin{eqnarray}
\label{u3}
\rho(r,t)= \frac{1}{1-t}\frac{{2}}{\pi}\frac{1}{\sqrt{1+2tx^2}},\nonumber\\
\times \left \lbrack 1-\frac{1-t}{t}\left (\sqrt{1+2tx^2}-1\right )\right\rbrack,
\end{eqnarray}
\begin{equation}
\label{u4}
x=\frac{r}{1-t}.
\end{equation}
For short times $t\rightarrow 0$, it is approximately given by
\begin{equation}
\label{u5}
\rho(r,t)\simeq \rho(r,0)+\frac{2}{\pi}\left (1-3r^2+\frac{3}{2}r^4\right )t,
\end{equation}
for $r\le r_{max}\simeq 1-t/4$ (see Fig. \ref{densiteTsmall}). For $0\le t\le t_{coll}=1$, the central density increases like (see Fig. \ref{rho0}):
\begin{equation}
\label{u6}
\rho(0,t)=\frac{2}{\pi}\frac{1}{1-t}.
\end{equation}
For $t\rightarrow t_{coll}=1$ and $r\rightarrow 0$, the density distribution takes the self-similar form
\begin{equation}
\label{u7}
\rho_{ss}(r,t)=\frac{1}{1-t}f\left (\frac{r}{1-t}\right ),
\end{equation}
with the invariant profile
\begin{equation}
\label{u8}
f(x)=\frac{2}{\pi}\frac{1}{\sqrt{1+2x^2}}.
\end{equation}
The central density corresponding to the self-similar solution (\ref{u7}) is
\begin{equation}
\rho_{ss}(0,t)=\frac{2}{\pi}\frac{1}{1-t},
\label{u8b}
\end{equation}
and it exactly coincides with Eq. (\ref{u6}). The singular profile corresponding to the self-similar solution (\ref{u7}) at $t=t_{coll}=1$ is
\begin{equation}
\label{u9}
\rho_{ss}(r,t_{coll})=\frac{\sqrt{2}}{\pi r}.
\end{equation}
The exact density profile at $t=t_{coll}=1$ is
\begin{equation}
\label{u10}
\rho(r,t_{coll})=\frac{\sqrt{2}}{\pi}\left (\frac{1}{r}-\sqrt{2}\right ).
\end{equation}
for $r\le r_{max}=1/\sqrt{2}$. The exact density profile is plotted in Fig. \ref{precollapseSS} using self-similar variables. This representation illustrate the fact that the solution becomes self-similar for $t\rightarrow t_{coll}=1$; indeed the curves tends to the invariant profile (\ref{u8}). The exact density profile is also plotted in Fig. \ref{densiteLL} in logarithmic variables. This representation illustrates the fact that the density displays a finite time singularity at $t\rightarrow t_{coll}=1$ and that, approaching the singularity, the profile becomes self-similar. For sufficiently small $t_{coll}-t$, the
 tail of the profile (for $r_0(t)\ll r\ll r_{max}(t)$) is well-approximated by the pure power law (\ref{u9}) but if we want to describe the distribution up to $r_{max}(t)\sim 1/\sqrt{2}$, we must use the truncated power law (\ref{u10}).

\begin{figure}[!h]
\begin{center}
\includegraphics[clip,scale=0.3]{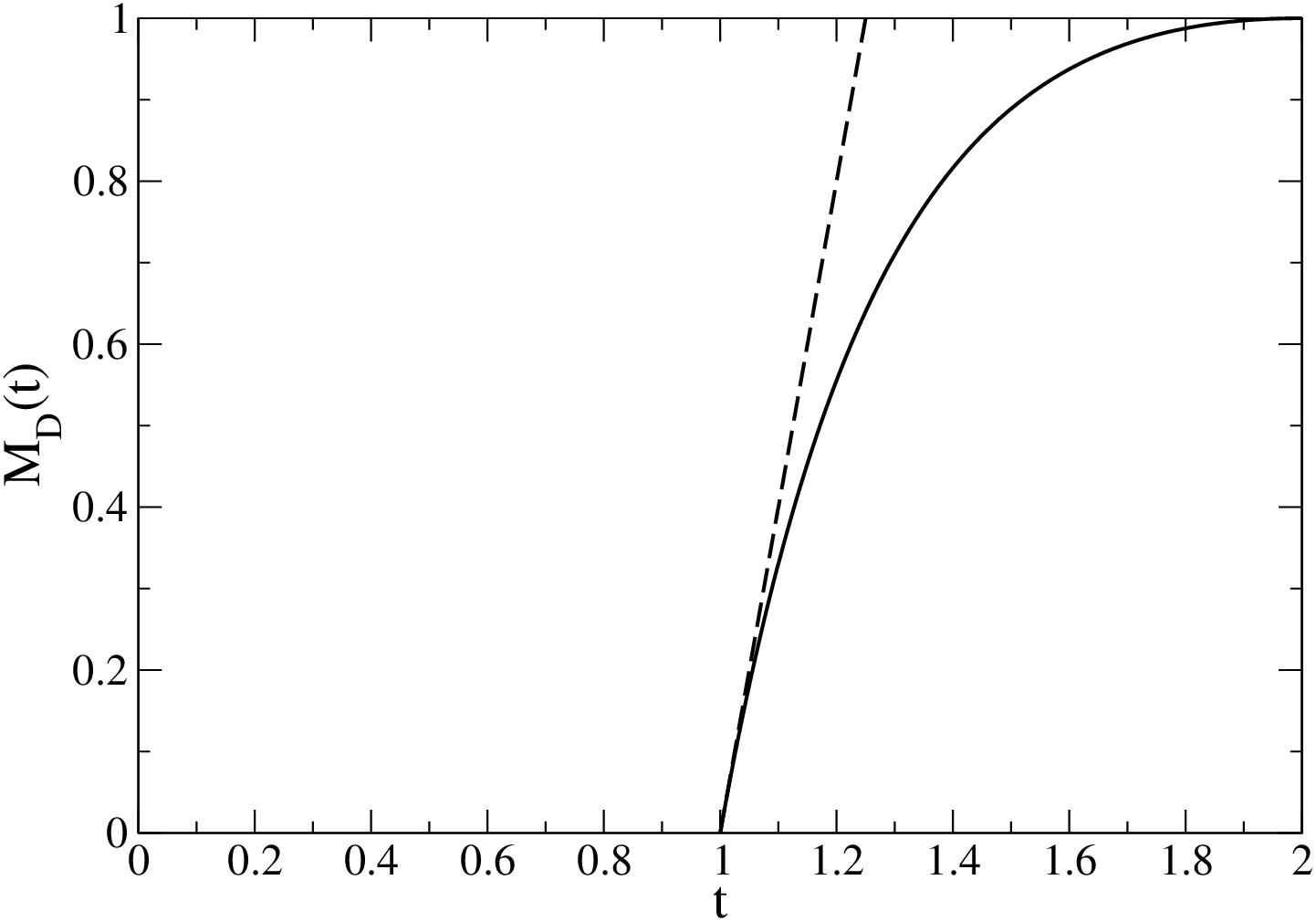}
\caption{Evolution of the mass contained in the Dirac peak. The dashed line corresponds to the self-similar solution. }
\label{MD}
\end{center}
\end{figure}

\begin{figure}[!h]
\begin{center}
\includegraphics[clip,scale=0.3]{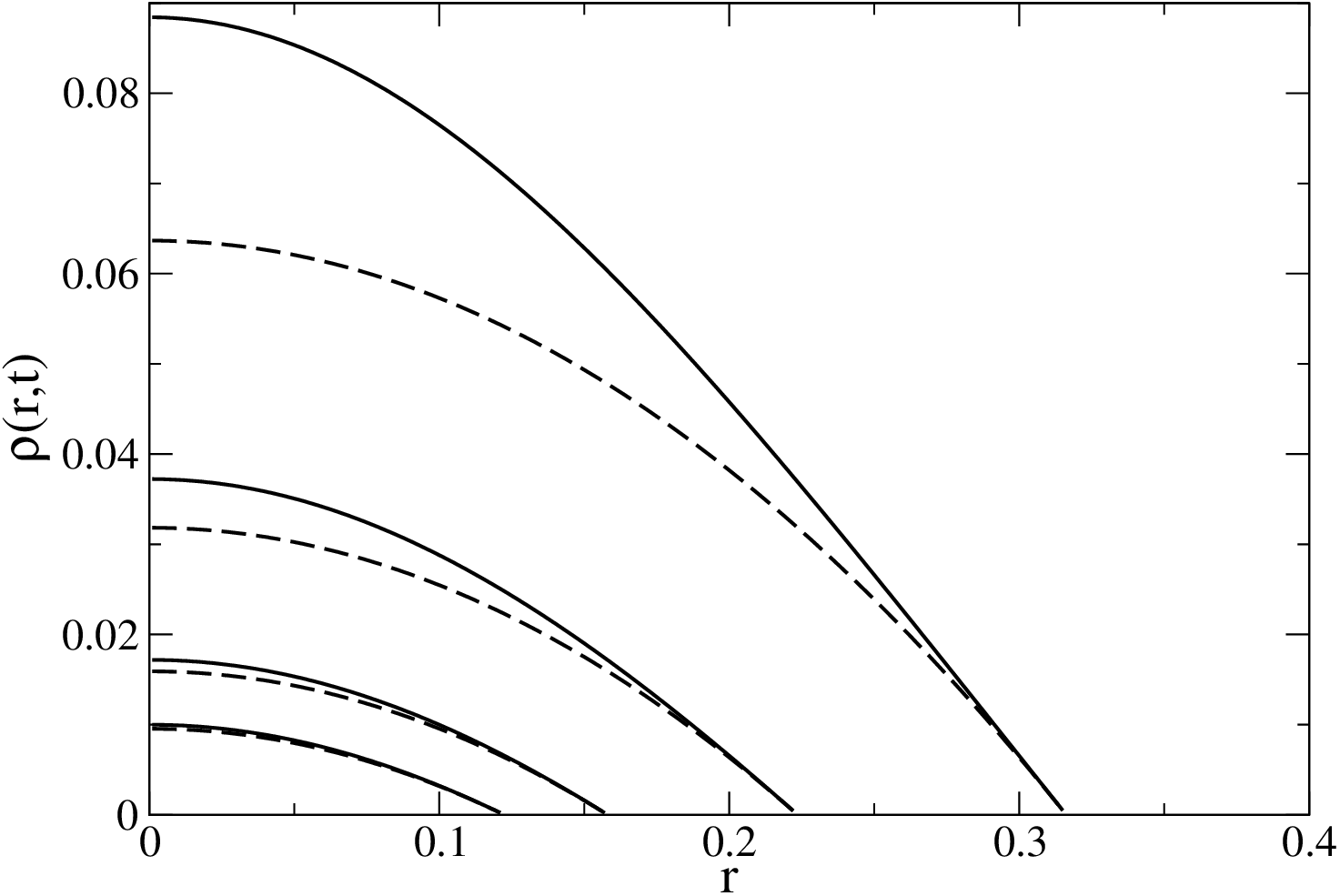}
\caption{Evolution of the density for $t\rightarrow t_{end}=2$. The full line corresponds to the exact profile and the dashed line to the approximation (\ref{u14}). We have represented $t=1.8, 1.9, 1.95, 1.97$ (top to bottom).}
\label{densiteTlarge}
\end{center}
\end{figure}

\begin{figure}[!h]
\begin{center}
\includegraphics[clip,scale=0.3]{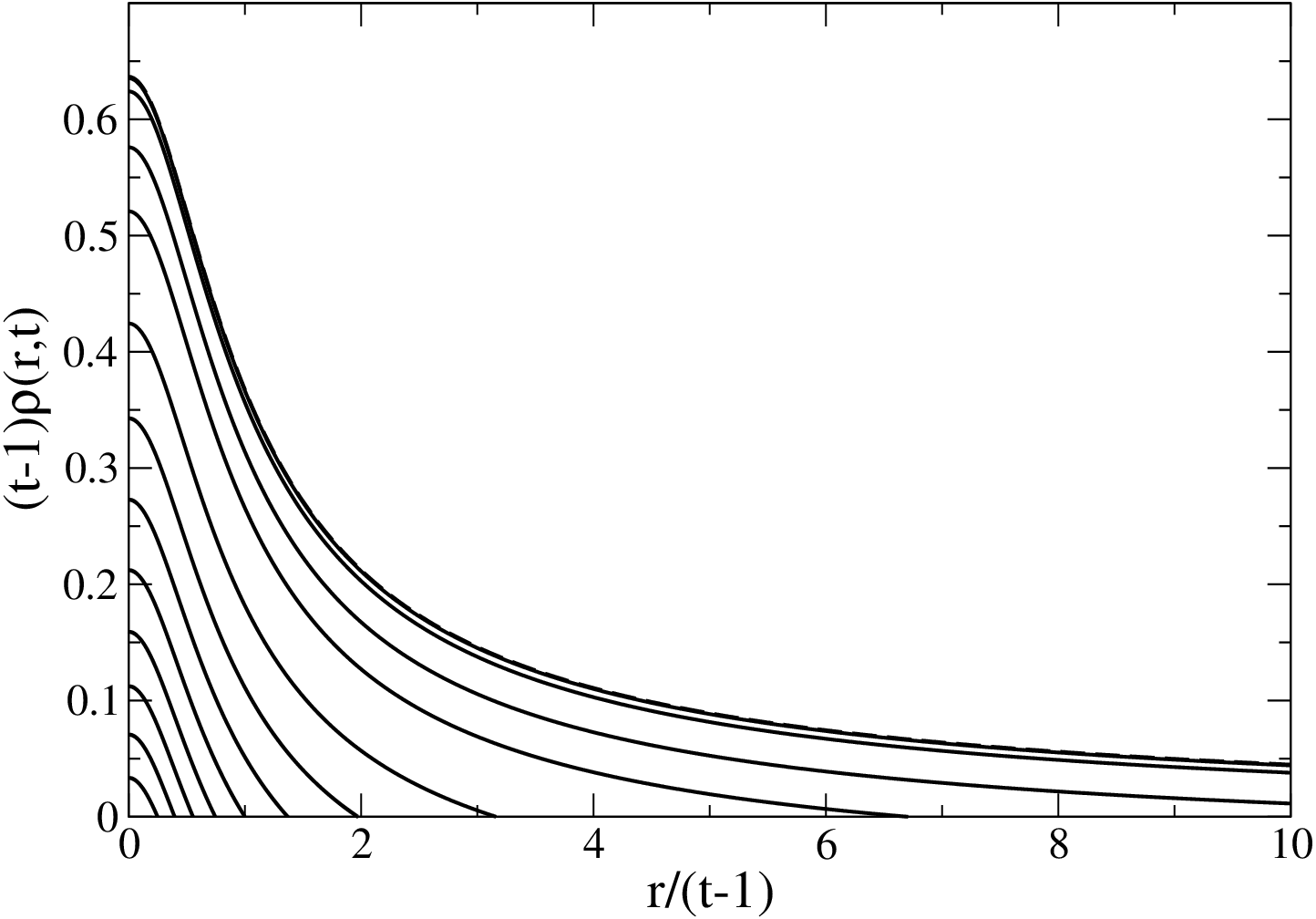}
\caption{Evolution of the density profile using self-similar variables. We have taken $t=1.001, 1.01, 1.05, 1.1, 1.2, 1.3, 1.4, 1.5, 1.6, 1.7, 1.8, 1.9, 2$. The full lines correspond to the exact profile (\ref{u12})-(\ref{u13}).  They tend to the invariant profile (\ref{u17}) represented as a dashed line.}
\label{postcollapseSS}
\end{center}
\end{figure}

\begin{figure}[!h]
\begin{center}
\includegraphics[clip,scale=0.3]{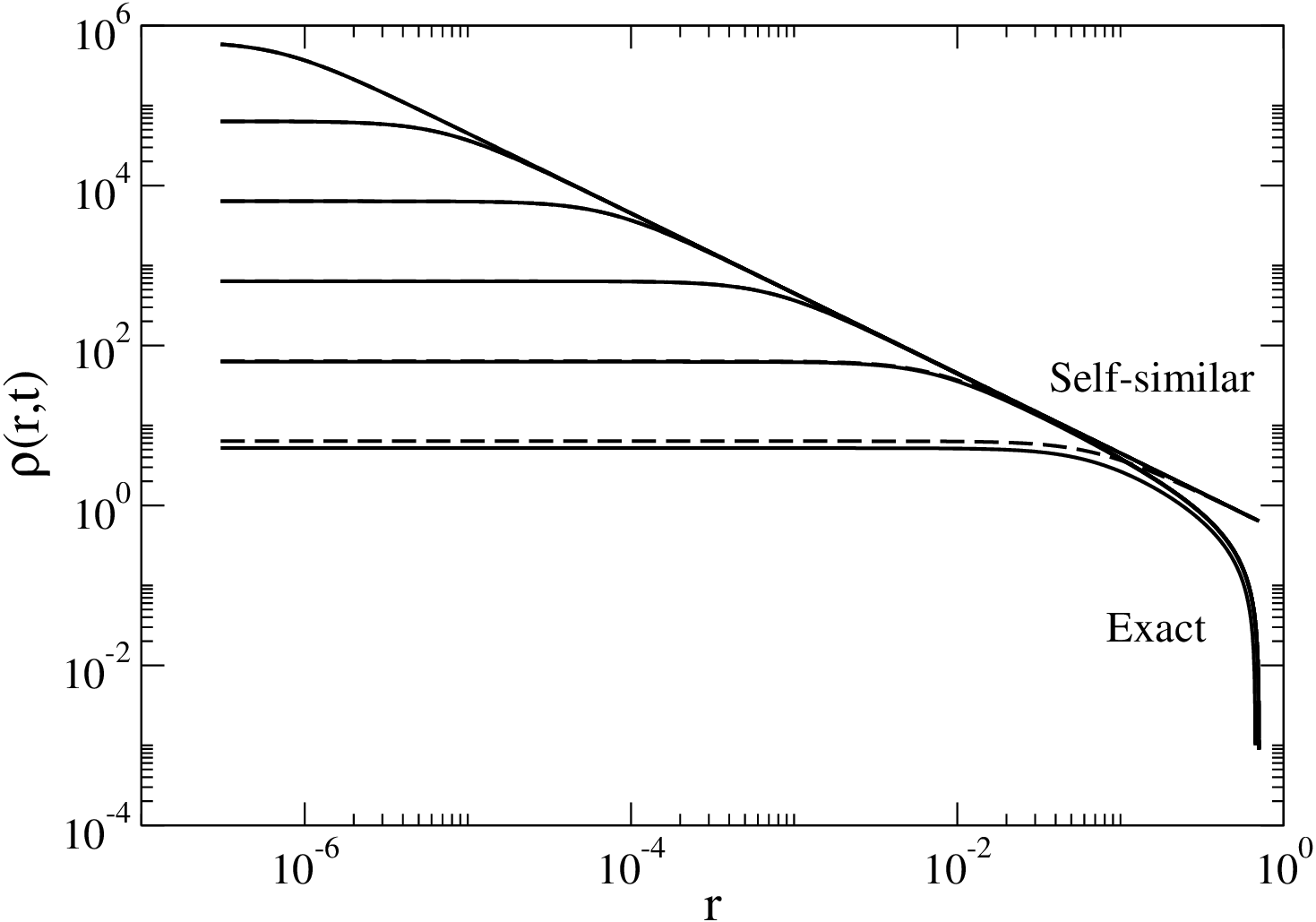}
\caption{Evolution of the density profile $\rho(r,t)$ in log-log plot. We have taken $t-t_{coll}=10^{-6}, 10^{-5}, 10^{-4}, 10^{-3}, 10^{-2}, 10^{-1}$. We have plotted the exact profile (\ref{u12})-(\ref{u13}) and the self-similar profile (\ref{u16})-(\ref{u17}) (dashed lines). At $t=t_{coll}$, the exact profile tend to the density (\ref{u10}) and the self-similar profile to the density (\ref{u9}). For sufficiently small $t-t_{coll}$, the self-similar profiles and the exact profiles are in good agreement except for $r\sim r_{max}(t)$. We start to see a difference in the central density for $t-t_{coll}=0.1$. This difference was imperceptible in the pre-collapse regime for the equivalent time $t_{coll}-t=0.9$. This is due to the difference in the laws of evolution of the central density (\ref{u6}), (\ref{u8b}) [identical] and (\ref{u15}),(\ref{u18}) [different] in the pre and post collapse regimes. }
\label{POSTdensiteLL}
\end{center}
\end{figure}

\begin{figure}[!h]
\begin{center}
\includegraphics[clip,scale=0.3]{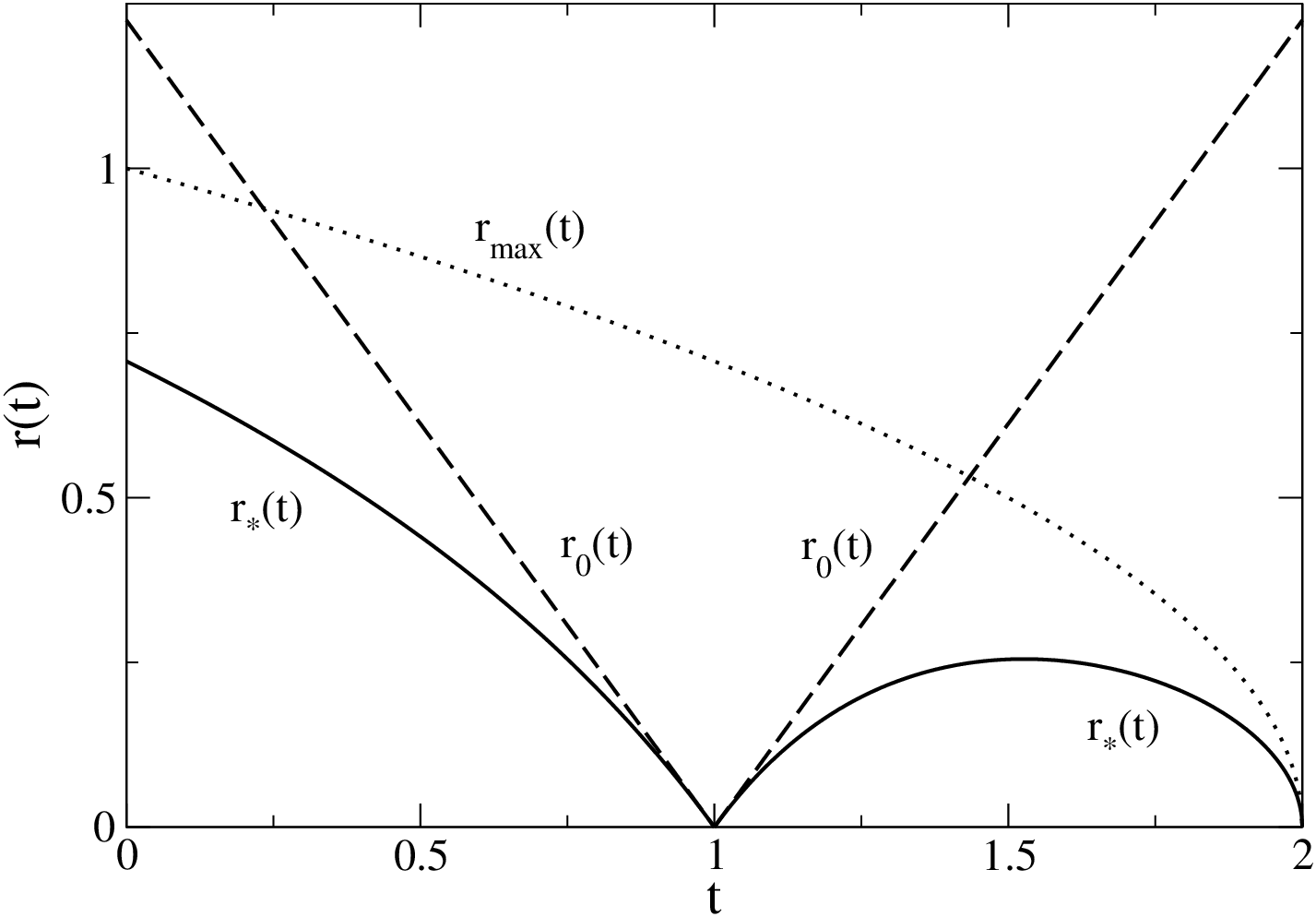}
\caption{Evolution of the radii $r_{max}$ (size of the system), $r_0$ (scaling variable) and $r_{*}$ (half-central density radius) with time.  }
\label{rstar}
\end{center}
\end{figure}

For $t_{coll}=1\le t\le t_{end}=2$ (post-collapse), the mass contained in the Dirac peak grows like (see Fig. \ref{MD}):
\begin{equation}
\label{u11}
M_D(t)=\frac{4}{t^2}(t-1).
\end{equation}
The exact residual density profile is
\begin{eqnarray}
\label{u12}
\rho(r,t)= \frac{1}{t-1}\frac{{2}}{\pi}\frac{1}{\sqrt{1+2tx^2}},\nonumber\\
\times \left \lbrack 1-\frac{t-1}{t}\left (1+\sqrt{1+2tx^2}\right )\right\rbrack,
\end{eqnarray}
\begin{equation}
\label{u13}
x=\frac{r}{t-1}.
\end{equation}
For $t\rightarrow t_{end}=2$, it is approximately given by
\begin{equation}
\label{u14}
\rho(r,t)=\frac{2}{\pi}\left (\frac{2-t}{2}-r^2\right ),
\end{equation}
for $r\le r_{max}\simeq \lbrack (2-t)/2\rbrack^{1/2}$ (see Fig. \ref{densiteTlarge}). For $t_{coll}=1\le t\le t_{end}=2$, the central density decreases like (see Fig. \ref{rho0}):
\begin{equation}
\label{u15}
\rho(0,t)=\frac{2}{\pi}\frac{1}{t-1}\left (\frac{2}{t}-1\right ).
\end{equation}
For $t\rightarrow t_{coll}=1^+$ and $r\rightarrow 0$, the residual density distribution takes the self-similar form
\begin{equation}
\label{u16}
\rho_{ss}(r,t)=\frac{1}{t-1}f\left (\frac{r}{t-1}\right ),
\end{equation}
with
\begin{equation}
\label{u17}
f(x)=\frac{2}{\pi}\frac{1}{\sqrt{1+2x^2}}.
\end{equation}
The central residual density corresponding to the self-similar solution (\ref{u16}) is
\begin{equation}
\rho_{ss}(0,t)=\frac{2}{\pi}\frac{1}{t-1}.
\label{u18}
\end{equation}
The singular profile corresponding to the self-similar solution (\ref{u16}) at $t=t_{coll}=1$ is given by Eq. (\ref{u9}) and the exact density profile at $t=t_{coll}=1$ is given by Eq. (\ref{u10}).
The exact density profile is plotted in Fig. \ref{postcollapseSS} using self-similar variables. This representation illustrates the fact that the solution is self-similar for $t\rightarrow t_{coll}=1^+$; indeed the curves tends to the invariant profile (\ref{u17}), the same as in the pre-collapse regime. Note, however, that the exact central density (\ref{u15}) is different from the self-similar central density (\ref{u18}) [they coincide only for $t\rightarrow t_{coll}=1^+$], contrary to the pre-collapse regime. This explains why the normalized central density $(t-1)\rho(0,t)$ is not a fixed point  in Fig. \ref{postcollapseSS} contrary to Fig. \ref{precollapseSS}. On the other hand, this figure shows the disappearance of the residual density profile at $t=t_{end}=2$ when all the mass has been absorbed in the Dirac peak at $r=0$. The exact density profile is also plotted in Fig. \ref{POSTdensiteLL} in logarithmic variables. This representation illustrates the singularity at $t=t_{coll}$ and the fact that the solution is self-similar for $t\rightarrow t_{coll}=1^+$. As in the pre-collapse regime, for sufficiently small $t-t_{coll}$, the tail the profile (for $r_0(t)\ll r\ll r_{max}(t)$) is well-approximated by the pure power law (\ref{u9}) but if we want to describe the distribution up to $r_{max}(t)\sim 1/\sqrt{2}$, we must use the truncated power law (\ref{u10}).

We represent in Fig. \ref{rstar} the evolution of the half central density radius.
For $0\le t\le t_{coll}=1$, it is given by
\begin{equation}
\label{f1z}
r_*(t)=\frac{1-t}{2-t}\left (2-\frac{t}{2}\right )^{1/2},
\end{equation}
and for $t_{coll}=1\le t\le t_{end}=2$, by
\begin{equation}
\label{f1a}
r_*(t)=\frac{1}{t}\left (\frac{1}{2}+\frac{1}{t}\right )^{1/2}(2-t)^{1/2}(t-1).
\end{equation}
For $t\rightarrow 0$, we have
\begin{equation}
\label{f1b}
r_*(t)\simeq \frac{1}{\sqrt{2}}\left (1-\frac{5t}{8}\right ),
\end{equation}
and for $t\rightarrow t_{end}=2$,
\begin{equation}
\label{f1c}
r_*(t)\simeq \frac{1}{{2}}(2-t)^{1/2}.
\end{equation}
For $t\rightarrow t_{coll}=1$, we find that
\begin{equation}
\label{f1d}
r_*(t)\simeq \left (\frac{3}{{2}}\right )^{1/2}|t-1|.
\end{equation}
In the pre-collapse regime, the half central density radius decreases
as the profile becomes more and more concentrated. In the
post-collapse regime, just after the collapse time $t_{coll}$, the half central density radius increases as the profile expands, then reaches a
maximum and finally decreases to zero as all the particles are
absorbed in the Dirac peak. In the self-similar regime, $r_{*}(t)$
has the same scaling as the core radius $r_0(t)$.

\begin{figure}[!h]
\begin{center}
\includegraphics[clip,scale=0.3]{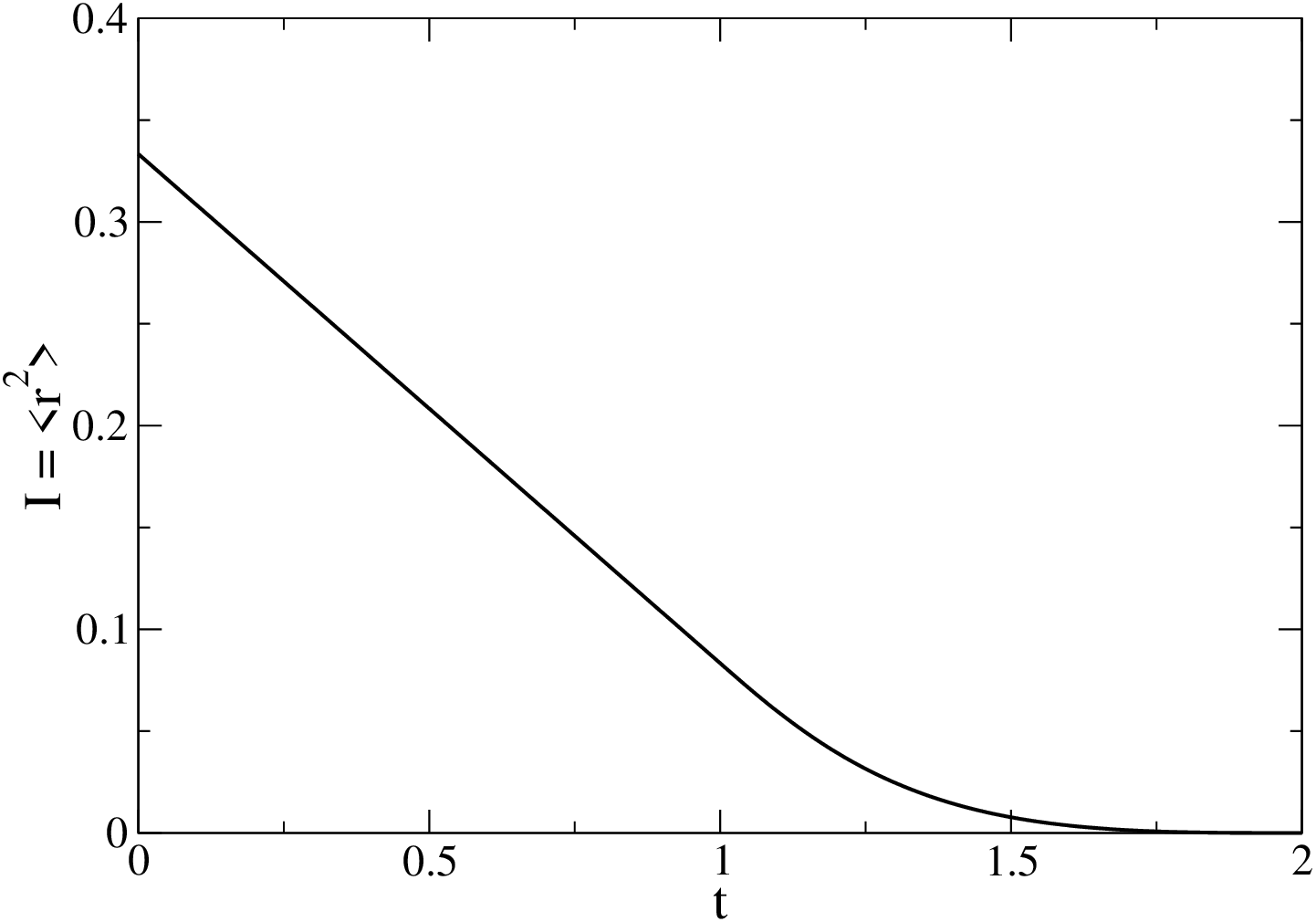}
\caption{Evolution of the moment of inertia $I=\langle r^2\rangle$ with time. }
\label{i}
\end{center}
\end{figure}

Finally, we represent in Fig. \ref{i} the evolution of the moment of inertia
\begin{equation}
\label{mi1}
I=\langle r^2\rangle=\int_{0}^{r_{max}(t)}\rho(r,t)r^2\, 2\pi rdr.
\end{equation}
It can be calculated from Eqs. (\ref{u3}) and (\ref{u12}) after lengthy computations or, more directly, from the virial theorem (\ref{vir1}) of Appendix \ref{sec_vir}. For $0\le t\le t_{coll}=1$, we find that
\begin{equation}
\label{mi2}
I(t)=-\frac{t}{4}-\frac{1}{3},
\end{equation}
and for  $t_{coll}=1\le t\le t_{end}=2$, we find that
\begin{equation}
\label{mi3}
I(t)=\frac{1}{12}\frac{1}{t^3}(2-t)^3(3t-2).
\end{equation}
Note that $I(t)=\langle r^2\rangle(t)$ is a monotonically decreasing function of time.

\section{Conclusion}

In this paper, we have provided an exact analytical solution
describing the gravitational collapse of a gas of self-gravitating
Brownian particles in the overdamped limit at $T=0$.  Starting from a
parabolic density profile, the system first develops a finite time
singularity at $t=t_{coll}$ in the pre-collapse regime, followed by
the formation and growth of a Dirac peak in the post-collapse
regime. Interestingly, our solution describes all the phases of the
dynamics. This extends the purely self-similar solution obtained
previously \cite{crs,sc,post} that is valid only close to
$t_{coll}$. Unfortunately, our method that exploits the deterministic
behavior of the system, is not valid anymore for $T>0$. In that case,
we must resort to other methods \cite{crs,sc,post,lushnikov} that are
much more complicated. However, the phenomenology of the collapse at
$T\neq 0$ (finite time singularity, growth of a Dirac peak,
self-similar solutions, pre and post collapse...) remains the same
(but, of course, the scaling exponents are different). Therefore, for
illustration of the general process of gravitational collapse of
self-gravitating Brownian particles, it is useful to have a
simple analytical solution such as the one described here.

It is interesting to compare the {\it isothermal collapse}
\cite{aaiso,crs} of self-gravitating Brownian particles in contact
with a heat bath to the collapse of isolated
stellar systems experiencing a {\it gravothermal catastrophe}
\cite{antonov,lbw}. These two types of systems
presents analogies and differences. For both systems, the pre-collapse is
self-similar and generates a finite time singularity where the central
density is infinite (see \cite{crs,sc} for self-gravitating Brownian
particles and \cite{henon,lbeggleton,cohn,lk} for stellar
systems). However, for stellar systems, the post-collapse regime leads
to a {\it binary star} surrounded by a hot halo \cite{henon} (statistical
equilibrium state in the microcanonical ensemble
\cite{lbw,paddy}) instead of a {\it Dirac peak} containing all the particles \cite{post} (statistical equilibrium state in the canonical ensemble
\cite{kiessling,aaiso}). The binary star can release sufficient
energy \cite{henon} to stop the collapse and even drive a re-expansion
of the system \cite{ilb}. Then, a series of gravothermal oscillations
should follow \cite{bs}. More references and discussions about the
statistical mechanics of self-gravitating systems in microcanonical
and canonical ensembles are given in the reviews
\cite{paddy,ijmpb}.

For certain systems, such as those discussed in the Introduction, it
is important to take into account both long-range and short-range
interactions. As we have seen, due to the attractive long-range
interaction, self-gravitating Brownian particles, bacterial
populations and colloids driven by attractive capillary interactions
can collapse. In that case, the central part of the system becomes
very dense. In the absence of short-range interactions, the collapse
generically leads to the formation of Dirac peaks. Of course, in
practice, Dirac peaks are unphysical and the density profile is
regularized by small-scale constraints. These small-scale constraints
can be due to finite size effects (the particles always have a finite
size and cannot interpenetrate), steric hindrance, short-range
interactions and, ultimately, quantum mechanics (Pauli exclusion
principle).  These interactions come into play when the system becomes
dense enough. Their effect is to provide a nonlinear pressure that can
halt the collapse and lead to a well-defined equilibrium state. An
example of this regularization is provided by a gas of
self-gravitating fermions in which gravitational collapse is balanced
by the pressure force arising from the Pauli exclusion principle
\cite{chandra,ijmpb,ribot}. Another example is provided by
chemotaxis \cite{degrad} where Dirac peaks are
replaced by smooth aggregates. In these examples, the Smoluchowski
equation at $T=0$ studied in this paper must be superseded by more general
equations of the form (\ref{i1}) where the pressure $p(\rho)$ prevents
complete collapse of the system. A microscopic justification of
these equations is given in \cite{gen,nfp,longshort} and some explicit
examples of short-range regularization are worked out in
\cite{ribot,degrad}.

Finally, we would like to point out that the numerous analogies
between self-gravitating Brownian particles, chemotaxis, colloids and
some nanosystems described in the Introduction lead to the fascinating
possibility of reproducing or mimicking a gravitational dynamics and a
gravitational collapse in the laboratory.  Experimental realizations
of these processes will certainly be developed in the future and represent an
interesting challenge.

\appendix

\section{Nonlinear mean field Fokker-Planck equations}
\label{sec_nfp}

The mean field drift-diffusion equation (\ref{i1}) is a subclass of the more general equation
\begin{equation}
\frac{\partial\rho}{\partial t}=\nabla\cdot (Dh(\rho)\nabla\rho+\chi g(\rho)\nabla\Phi),
\label{nfp1}
\end{equation}
coupled to Eq. (\ref{i2}) or to a potential $\Phi=\rho *  u$ (where $u({\bf r},{\bf r}')$ is a binary potential of interaction and $*$ denotes the convolution product) introduced in \cite{gen,nfp}. Here, both the diffusion coefficient $Dh(\rho)$ and the mobility $\chi g(\rho)/\rho$ can depend on the density. The fact that the diffusion and the mobility depend on the density in complex systems is not surprising. For example, in a dense fluid, we understand that the motion of a given particle is hampered by interactions with its neighbors so that its mobility is reduced and its diffusivity modified (with respect to a dilute medium). For long-range interactions, these equations generalize the standard Debye-H\"uckel, Smoluchowski-Poisson and Keller-Segel models. For short-range interactions, we can make the gradient expansion  $\Phi\simeq -a\rho-\frac{b}{2}\Delta\rho$ with $a=-S_d\int_0^{+\infty}u(q)q^{d-1}\, dq$ and $b=-\frac{1}{d}S_d\int_0^{+\infty}u(q)q^{d+1}\, dq$ \cite{et}, and we obtain a generalization of the Cahn-Hilliard equations. Equation (\ref{nfp1}) was introduced in \cite{gen,nfp} as a nonlinear  mean field Fokker-Planck equation associated with a generalized thermodynamical formalism. Later, Holm \& Putkaradze \cite{holm} considered a very related model
\begin{equation}
\frac{\partial\rho}{\partial t}=\nabla\cdot (D\nabla\overline{\rho}+\mu(\overline{\rho})\rho\nabla\Phi),
\end{equation}
where $\Phi=\rho *  u$ and $\overline{\rho}=\rho *  H$, where $u(|{\bf r}-{\bf r}'|)$ is a binary potential and $H(|{\bf r}-{\bf r}'|)$  a local filter function. It differs from Eq. (\ref{nfp1}) only in the replacement of $\rho$ by $\overline{\rho}$ in the diffusion and mobility terms.

\section{Virial theorem in $d=2$}
\label{sec_vir}

In Appendix H of Ref. \cite{cm}, one of the authors has obtained the proper form of the virial theorem associated with the 2D Smoluchowski-Poisson system which is valid {\it both} in the pre and post collapse regimes. Its general expression is given by
\begin{equation}
\label{vir1}
\frac{1}{4}\xi\frac{dI}{dt}=M(t)\left\lbrack \frac{k_B T}{m}-\frac{GM_D(t)}{2}-\frac{GM(t)}{4}\right \rbrack,
\end{equation}
where $I=\int\rho r^2\, d{\bf r}=M\langle r^2\rangle$ is the moment of inertia, $M_{D}(t)$ the mass contained in the Dirac peak and $M(t)=M-M_{D}(t)$ the residual mass contained in the regular density profile. In the absence of Dirac peak, Eq. (\ref{vir1}) reduces to the standard expression \cite{virial1}:
\begin{equation}
\label{vir2}
\frac{1}{4}\xi\frac{dI}{dt}=Nk_B(T-T_c),
\end{equation}
involving the critical temperature
\begin{equation}
\label{vir3}
k_BT_c=\frac{GMm}{4}.
\end{equation}
This equation is {\it closed} and can be integrated into
\begin{equation}
\label{vir4}
I(t)=\frac{4Nk_B}{\xi}(T-T_c)t+I_{0}.
\end{equation}
This is valid for any time when $T\ge T_c$ and in the pre-collapse regime when $T< T_c$. In the post-collapse regime when $T<T_c$, we must come back to the general expression (\ref{vir1}). However, this equation is not closed since it involves the mass $M_{D}(t)$ contained in the Dirac peak. Introducing the mean square displacement $\langle r^2\rangle=I(t)/M$, Eq. (\ref{vir4}) shows that, for $T\ge T_c$, the motion of a particle is diffusive with an effective diffusion coefficient \cite{virial1}:
\begin{eqnarray}
\label{vir4b}
D(T)=\frac{k_BT}{\xi m}\left (1-\frac{T_c}{T}\right ),
\end{eqnarray}
that is independent on the initial condition.

At $T=0$, Eq. (\ref{vir1}) reduces to the form
\begin{equation}
\label{vir5}
\xi\frac{dI}{dt}=-G\left\lbrack M^2-M_{D}(t)^2\right \rbrack.
\end{equation}
In the pre-collapse regime, $t\le t_{coll}$, we obtain
\begin{equation}
\label{vir6}
I(t)=-\frac{GM^2}{\xi}t+I_0.
\end{equation}
In the post-collapse regime, $t_{coll}\le t\le t_{end}$, we have established in Sec. \ref{sec_post} that the mass of the Dirac peak increases like (in $d=2$):
\begin{equation}
\label{vir7}
M_D(t)=4M\left (\frac{t_{coll}}{t}\right )^2\left (\frac{t}{t_{coll}}-1\right ).
\end{equation}
Substituting this expression in Eq. (\ref{vir5}) and integrating Eq. (\ref{vir5}) between $t$ and $t_{end}$ at which $I(t_{end})=0$, we find that
\begin{equation}
\label{vir8}
\frac{I(t)}{MR^2}=\frac{1}{12}\left (\frac{t_{coll}}{t}\right )^3\left (2-\frac{t}{t_{coll}}\right )^3\left (\frac{3t}{t_{coll}}-2\right ).
\end{equation}
We have checked, after lengthy calculations, that Eqs. (\ref{vir6}) and (\ref{vir8}) can also be obtained directly from the exact analytical density profiles (\ref{d1b}) and (\ref{da2}). The evolution of the moment of inertia (proportional to the dispersion of the particles) is plotted in Fig. \ref{i}.

There is no closed expression of the virial theorem when $d\neq 2$.
Therefore, the moment of inertia must be calculated
directly from the exact analytical expressions of the density profile
given in the main part of the paper. We shall give only particular
values of the moment of inertia in $d$ dimensions
\begin{equation}
I(t)=\int_{0}^{r_{max}(t)}\rho(r,t)r^{2}\, S_d r^{d-1}dr.
\end{equation}
For $t\rightarrow 0$, using Eqs. (\ref{p21}) and (\ref{p22}), we find that
\begin{equation}
\label{izero}
I(t)\simeq I_0-\frac{12}{(d+4)(d+6)}MR^2 \frac{t}{t_{coll}}.
\end{equation}
For $t=t_{coll}$, using Eqs. (\ref{px2}) and (\ref{px3}), we get
\begin{equation}
\label{icoll}
I(t_{coll})=\frac{d^3}{d+2}\left (\frac{d}{d+2}\right )^{2/d}\frac{1}{d^2+2d+4}MR^2.
\end{equation}
For $t\rightarrow t_{end}$, using Eqs. (\ref{a2}) and (\ref{a3}), we obtain
\begin{equation}
\label{iend}
I(t)\sim \frac{d(d+2)}{2(d^2+3d+2)}MR^2 \left (\frac{t_{end}-t}{t_{end}}\right )^{2(d+1)/d}.
\end{equation}
Using Eqs. (\ref{a3}) and (\ref{a2c}), we check that this last
expression can be written
\begin{equation}
\label{iendb}
I(t)\sim \frac{d^2}{d^2+3d+2}M(t)r_{max}(t)^2.
\end{equation}
For $d=2$, the expressions (\ref{izero}), (\ref{icoll}) and
(\ref{iend}) are consistent with the general results (\ref{vir6}) and
(\ref{vir8}) valid for all times. For the uniform sphere in $d$ dimension, using Eqs. (\ref{h4})-(\ref{h6}), we obtain
\begin{eqnarray}
\label{iunif}
I(t)=\frac{d}{d+2}MR^2\left (1-\frac{dGM}{\xi R^d}t\right )^{2/d}=\frac{d}{d+2}MR(t)^2.\nonumber\\
\end{eqnarray}
In $d=2$, this expression reduces to Eq. (\ref{vir6}) as it should.

\section{Derivation of the generalized mean field Smoluchowski equation}
\label{sec_deriv}

In this Appendix, following \cite{longshort}, we provide a new
derivation of the generalized mean field Smoluchowski equation
(\ref{i1}). In previous works
\cite{gen,nfp,cll}, this equation was derived from a notion of
generalized thermodynamics. In that case, the nonlinear pressure was
due to a bias in the transition probabilities leading to
non-Boltzmannian distributions \cite{kaniadakis}. Here, we show that
the same equation can be derived from the Dynamic Density Functional
Theory (DDFT) used in the theory of simple liquids \cite{marconi}. In
that case, the nonlinear pressure is due to the correlations induced
by the short-range interactions. Although physically distinct, these
processes lead to the same type of macroscopic equations. A more
detailed discussion, as well as the derivation of more general kinetic
equations, is given in
\cite{longshort}.

We consider the overdamped dynamics of $N$ Brownian particles in interaction
governed by the coupled stochastic equations \cite{hb2}:
\begin{equation}
\label{o1} \frac{d{\bf r}_{i}}{dt}=-\mu\nabla_{i}U({\bf
r}_{1},...,{\bf r}_{N})+\sqrt{2D}{\bf R}_{i}(t),
\end{equation}
where $U({\bf r}_1,...,{\bf r}_N)=m^2\sum_{i<j}u(|{\bf r}_i-{\bf
r}_j|)$ is the potential of interaction and ${\bf R}_{i}(t)$ is a
Gaussian white noise such that $\langle {\bf R}_{i}(t)\rangle={\bf 0}$
and $\langle
R_{i}^{\alpha}(t)R_{j}^{\beta}(t')\rangle=\delta_{ij}\delta_{\alpha\beta}\delta(t-t')$. Here,
$i=1,...,N$ label the particles and $\alpha=1,...,d$ the coordinates
of space.  The diffusion coefficient $D$ is related to the mobility
$\mu=1/(\xi m)$ and the temperature $T$ by the Einstein relation
$D=\mu k_B T$ \cite{risken}. The time evolution of the $N$-body distribution
$P_{N}({\bf r}_1,...,{\bf r}_N,t)$ is governed by the $N$-body
Fokker-Planck equation
\begin{equation}
\label{o3} \xi{\partial P_{N}\over\partial t}=\sum_{i=1}^{N}
{\partial\over\partial {\bf r}_{i}}\cdot \biggl\lbrack
\frac{k_B T}{m}{\partial P_{N}\over\partial {\bf r}_{i}}+\frac{1}{m}
P_{N}{\partial\over\partial {\bf r}_{i}}U({\bf r}_{1},...,{\bf
r}_{N})\biggr\rbrack.
\end{equation}
This particular Fokker-Planck equation is called the $N$-body Smoluchowski equation. Its steady state is the   Gibbs canonical distribution $P_N=\frac{1}{Z}e^{-\beta U}$.

It is easy to derive from Eq. (\ref{o3}) the equivalent of the BBGKY
hierarchy for the reduced distribution functions
\cite{hb2}. Introducing the local density $\rho({\bf
r},t)=NmP_1({\bf r},t)$ and the two-body distribution function
$\rho_2({\bf r},{\bf r}',t)=N(N-1)m^2P_2({\bf r},{\bf r}',t)$, the
first equation of the BBGKY-like hierarchy is the exact Smoluchowski
equation
\begin{eqnarray}
\label{o7}  \xi\frac{\partial\rho}{\partial t}=\nabla\cdot \left\lbrack
 \frac{k_B T}{m} \nabla\rho+ \int \rho_2({\bf r},{\bf r}',t)\nabla u(|{\bf r}-{\bf r}'|)\, d{\bf r}'\right \rbrack,\nonumber\\
\end{eqnarray}
where we have used the fact that the particles are identical. This
equation is not closed since it involves the two-body correlation
function $\rho_2({\bf r},{\bf r}',t)$. We must therefore introduce
some approximations to evaluate this term. We shall assume that the
potential of interaction $u=u_{LR}+u_{SR}$ is the sum of a long-range
potential $u_{LR}$ and a short-range potential $u_{SR}$. For systems
with long-range interactions, it is known that the mean field
approximation is exact in a proper thermodynamic limit $N\rightarrow
+\infty$ \cite{messer,cdr}. Therefore, concerning the long-range
potential $u_{LR}$, we shall make the approximation $\rho_{2}({\bf
r},{\bf r}',t)=\rho({\bf r},t)\rho({\bf r}',t)$ leading to
\begin{equation}
\label{o10c}
\int \rho_2({\bf r},{\bf r}',t)\nabla u_{LR}(|{\bf r}-{\bf r}'|)\, d{\bf r}'=\rho({\bf r},t)\nabla\Phi({\bf r},t),
\end{equation}
with
\begin{equation}
\label{o10} \Phi({\bf r},t)=\int \rho({\bf r}',t)u_{LR}(|{\bf r}-{\bf r}'|)\, d{\bf r}'.
\end{equation}
To evaluate the integral corresponding to the short-range interactions, we shall use an approximation that has become standard in the dynamic density functional theory (DDFT) of fluids  \cite{marconi} and take
\begin{eqnarray}
\label{o12} \int \rho_{2}({\bf r},{\bf r}',t)\nabla u_{SR}(|{\bf r}-{\bf r}'|)\, d{\bf r}'\simeq \rho({\bf r},t)\nabla \frac{\delta F_{ex}}{\delta\rho}[\rho({\bf r},t)],\nonumber\\
\end{eqnarray}
where $F_{ex}[\rho]$ is the excess free energy \footnote{The excess
free energy $F_{ex}[\rho]$ is a non-trivial functional determined by
the short-range interactions. All the difficulty in the theory of
fluids is to find some approximate forms of this functional. Once this
functional is determined, the density profile, as well as all the
$n$-point correlation functions, can be obtained via functional
differentiation. Inversely, the excess free energy is often obtained
from the study of the correlation functions. The excess free energy
$F_{ex}$ is known exactly only in a few particular cases, but very
good approximations can be devised in more general cases
\cite{evans,hansen}.} calculated at equilibrium. This relation is
exact at equilibrium \cite{evans} and the approximation consists in
extending it out-of-equilibrium with the actual density $\rho({\bf
r},t)$ calculated at each time. This closure is equivalent to assuming
that the two-body dynamic correlations are the same as those in an
equilibrium fluid with the same one body density profile.  Although it
is not possible to ascertain the validity of this approximation in the
general case, it has been observed for the systems considered that
this approximation gives remarkable agreement with direct Brownian
$N$-body simulations. With the approximations (\ref{o10c}) and
(\ref{o12}), Eq. (\ref{o7}) becomes
\begin{equation}
\label{o13}  \xi\frac{\partial\rho}{\partial t}=\nabla\cdot \left\lbrack
 \frac{k_B T}{m}\nabla\rho+\rho\nabla \frac{\delta F_{ex}}{\delta\rho}+\rho\nabla\Phi\right \rbrack,
\end{equation}
which is closed. The total free energy is
\begin{equation}
\label{ftot} {F}[\rho]=\frac{1}{2}\int \rho\Phi\, d{\bf r}+k_B T\int {\rho\over m}\ln {\rho\over m}d{\bf r}+F_{ex}[\rho].
\end{equation}
The steady state of Eq. (\ref{o13}) minimizes (\ref{ftot}) at fixed mass and is given by $\delta F-\mu\delta M=0$ where $\mu$ is a Lagrange multiplier. This yields $\delta F/\delta\rho=\mu$ i.e.
\begin{equation}
\label{e36vf}
\rho({\bf r})=A e^{-\beta m \left (\Phi+\frac{\delta F_{ex}}{\delta\rho}\right )}.
\end{equation}

In a fluid, the local pressure is of the form $p=p(\rho,T)$. Since the
temperature $T$ is fixed in the case of Brownian particles (canonical
description), the pressure is barotropic and we shall simply write
$p=p(\rho)$. In principle, the excess free energy $F_{ex}[\rho]$ can
depend on the gradients of the density. This is particularly important
for a fluid close to an interface \cite{evans}. Here, we shall assume
that the density varies on a distance that is large with respect to
the range of intermolecular forces. This is the case if the density
distribution is mainly due to long-range interactions, as we shall
assume in the following. With this assumption, the free energy is of
the form \cite{longshort}:
\begin{equation}
\label{e43} {F}[\rho]=\frac{1}{2}\int \rho\Phi\, d{\bf r}+\int \rho\int^{\rho}\frac{p(\rho_1)}{\rho_1^2}\, d\rho_1\, d{\bf r}.
\end{equation}
The excess free energy is therefore
\begin{equation}
\label{e44} {F}_{ex}[\rho]=\int \rho\int^{\rho}\frac{p(\rho_1)}{\rho_1^2}\, d\rho_1\, d{\bf r}-k_B T\int {\rho\over m}\ln {\rho\over m}d{\bf r}.
\end{equation}
We note the relation
\begin{eqnarray}
\label{e45}
\nabla p(\rho)=\frac{k_{B}T}{m}\nabla\rho+\rho\nabla\frac{\delta F_{ex}}{\delta\rho}=\nabla p_{id}+\nabla p_{ex},
\end{eqnarray}
where $p_{id}({\bf r})=\rho({\bf r}) k_B T/m$ is the ideal gas law and
$p_{ex}$ the excess pressure due to short-range interactions
\cite{hansen}. This relation can be used to determine the equation of
state $p(\rho)$ corresponding to the excess free energy $F_{ex}[\rho]$
and {\it vice versa}. For an ideal fluid ($F_{ex}=0)$, we recover the
perfect gas law $p({\bf r})=\rho({\bf r}) k_B T/m$. For a
free energy of the form (\ref{e43}), using identity (\ref{e45}),
Eq. (\ref{o13}) can be rewritten
\begin{equation}
\label{o14}  \xi\frac{\partial\rho}{\partial t}=\nabla\cdot (
 \nabla p+\rho\nabla\Phi ).
\end{equation}
The  equilibrium state is given by the condition of hydrostatic equilibrium
\begin{equation}
\label{o14b}
\nabla p+\rho\nabla\Phi={\bf 0}.
\end{equation}
Equation (\ref{o14}) is a generalized mean field Smoluchowski equation
including a generically nonlinear barotropic pressure $p(\rho)$ due to
short-range interactions and a mean field potential $\Phi=u_{LR} *
\rho$ due to long-range interactions. As we have indicated in the
Introduction, this equation arises in several physical problems such
as self-gravitating Brownian particles \cite{grossmann}, chemotaxis
\cite{ks}, colloids with capillary interactions \cite{colloids},
etc. Combining results issued from the physics of systems with
long-range interactions \cite{cdr} and from the dynamic density
functional theory of fluids \cite{marconi}, this equation has been
justified from a microscopic model \cite{longshort}.

The generalized mean field Smoluchowski equation (\ref{o13}) can be written in terms of the free energy functional (\ref{ftot}) as
\begin{eqnarray}
\label{o15}
\frac{\partial\rho}{\partial t}=\nabla\cdot \left( \frac{1}{\xi}\rho\nabla\frac{\delta {F}}{\delta\rho}\right).
\end{eqnarray}
This equation monotonically decreases the free energy functional (\ref{ftot}) which plays therefore  the role of a Lyapunov functional. Indeed, a straightforward calculation leads to the  $H$-theorem appropriate to the canonical ensemble
\begin{eqnarray}
\label{o16}
\dot {F}=\int \frac{\delta F}{\delta\rho}\frac{\partial\rho}{\partial t}d{\bf r}=\int \frac{\delta F}{\delta\rho}\nabla\cdot \left ( \frac{1}{\xi}\rho\nabla\frac{\delta F}{\delta\rho}\right ) d{\bf r}\nonumber\\
=-\int\frac{1}{\xi}\rho \left (\nabla\frac{\delta F}{\delta\rho}\right )^{2} d{\bf r}\le 0.
\end{eqnarray}
For a steady state, $\dot F=0$, the last term in parenthesis must vanish so that $\delta F/\delta\rho$ is uniform. This leads to Eq. (\ref{e36vf}). Therefore, a density $\rho({\bf r})$ is a steady state of the generalized mean field Smoluchowski  equation (\ref{o15}) iff it is a critical point of $F$ at fixed mass. Furthermore, it can be shown that a steady state is linearly dynamically stable with respect to the generalized Smoluchowski equation (\ref{o15}) iff it is a (local) minimum of $F$ at fixed mass \cite{frank,nfp}. This is consistent with the condition of thermodynamical equilibrium. If $F$ is bounded from below \footnote{This is not always the case. For example, the free energy associated with the Smoluchowski-Poisson system describing self-gravitating Brownian particles is not bounded from below \cite{aaiso,crs,kiessling,sc}. In that case, the system can experience an isothermal collapse. However, there also exists long-lived metastable states on which the system can settle \cite{aaiso,crs,metastable}.}, we know from Lyapunov's direct method that the system will converge towards a  (local) minimum of $F$ at fixed mass $M$ for $t\rightarrow +\infty$. If several (local) minima exist (metastable states), the choice of the selected equilibrium will depend on a complicated notion of basin of attraction.

{\it Remark:} note that Eq. (\ref{o15}) can be justified  in a phenomenological manner from the linear thermodynamics of Onsager if we interpret it as a continuity equation $\partial_t\rho+\nabla\cdot {\bf J}=0$ with a current ${\bf J}=-(1/\xi)\nabla {\delta {F}}/{\delta\rho}$ proportional to the gradient of a potential $\mu({\bf r})={\delta {F}}/{\delta\rho}$ that is uniform at equilibrium (see Eq. (\ref{e36vf})). This is precisely the way in which this equation was introduced in the physics of liquids \cite{evans} and, more generally, in \cite{frank,nfp}.

\section{The formation of a peripheric Dirac peak}
\label{sec_peri}

In Sec. \ref{sec_formal} we have mentioned  that the evolution of the system is qualitatively different whether $M(a,0)/a^d$ is an increasing or a decreasing function of $a$. In the main part of the paper, we have considered the physical situation where $M(a,0)/a^d$ decreases. We have shown that such initial conditions lead to the growth of a Dirac peak at the origin. In this Appendix, we illustrate on a specific example the situation where $M(a,0)/a^d$ increases. We show that such initial conditions lead to the formation of a peripheric Dirac peak (a $d$-dimensional annulus) that progressively converges towards the center of the domain by absorbing the interior particles. To simplify the formulae, we choose a system of units such that $M=\xi=G=1$. We also assume that the initial size of the system is $R=1$.

\begin{figure}[!h]
\begin{center}
\includegraphics[clip,scale=0.3]{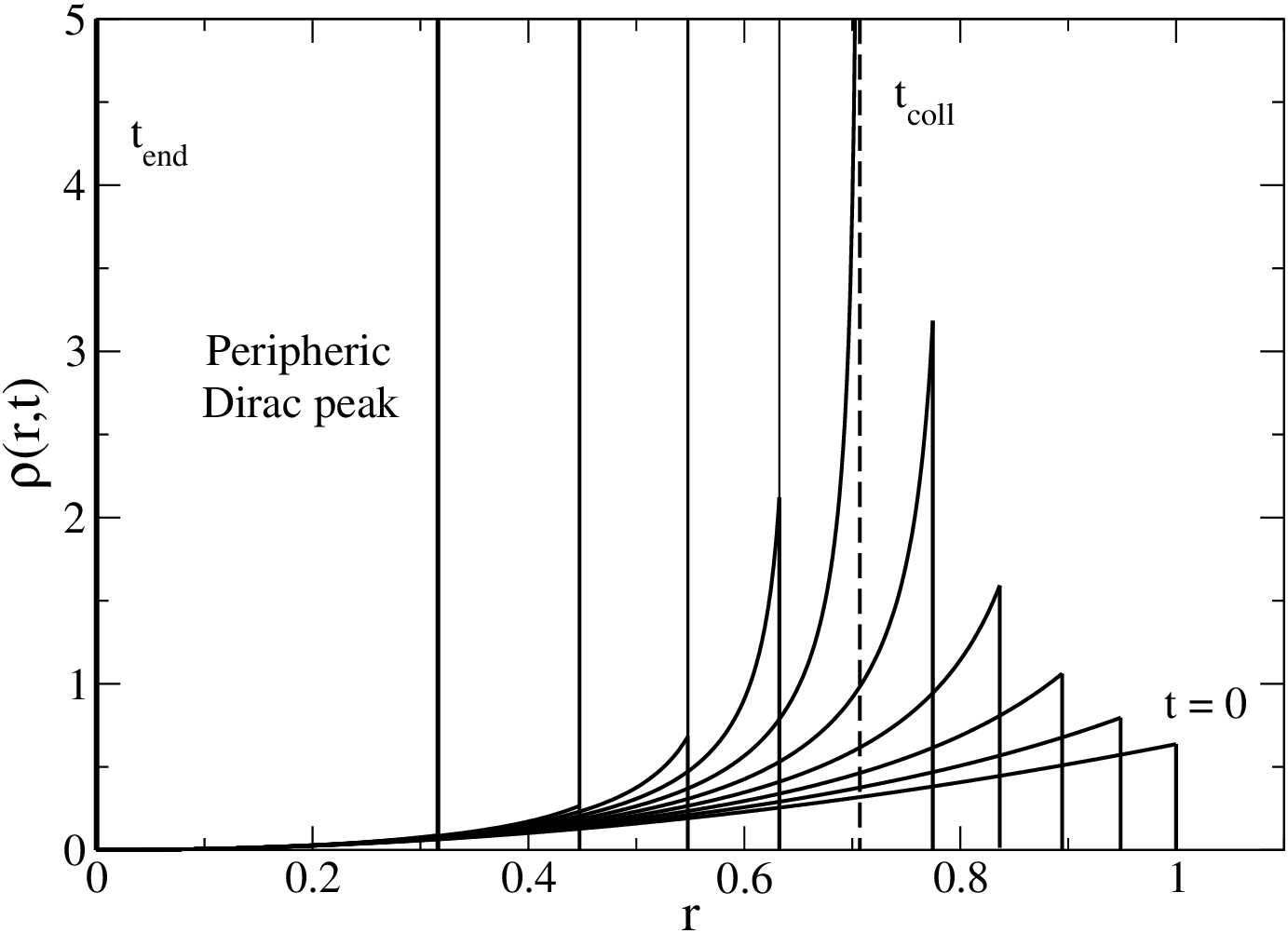}
\caption{Evolution of the density profile $\rho(r,t)$ in $d=2$. The time $t$ goes from $0$ to $t_{end}=0.5$ and is represented every $0.05$ time steps. The vertical line corresponds to the peripheric Dirac peak, appearing at $t=t_{coll}=1/4$ and containing all the mass at $t=t_{end}=1/2$. }
\label{peripheric}
\end{center}
\end{figure}

An analytical solution can be obtained by considering an initial condition of the form
\begin{eqnarray}
\label{peri1}
M(a,0)=a^{2d},
\end{eqnarray}
for $a\le 1$ and $M(a,0)=1$ for $a\ge 1$. The corresponding density profile, given by Eq. (\ref{f8}), is
\begin{eqnarray}
\label{peri2}
\rho(a,0)=\frac{2d}{S_d}a^{d},
\end{eqnarray}
for $a\le 1$ and $\rho(a,0)=0$ for $a\ge 1$. The density increases with the radius  for $a\le 1$ and vanishes discontinuously at $a=1$. According to Eq. (\ref{f11}), the position at time $t$ of the particle initially located at $a$ is
\begin{eqnarray}
\label{peri3}
r^d=a^d-d a^{2d}t.
\end{eqnarray}
According to Eqs. (\ref{f9}) and (\ref{peri1}), the mass profile at time $t$ is
\begin{eqnarray}
\label{peri4}
M(r,t)=a^{2d},
\end{eqnarray}
where $a$ is related to $r$ and $t$ through Eq. (\ref{peri3}). The size of the system at time $t$, i.e. the position of the last particle initially located at $a=1$, is
\begin{eqnarray}
\label{peri5}
r_{max}(t)=(1-d t)^{1/d}.
\end{eqnarray}
From Eq. (\ref{peri5}), all the particles have collapsed at $r=0$ at the final time
\begin{eqnarray}
\label{peri7}
t_{end}=\frac{1}{d}.
\end{eqnarray}
Solving for $a$ in Eq. (\ref{peri3}), which is a second degree equation in $a^d$,  and substituting the resulting expression in Eq. (\ref{peri4}), we obtain
\begin{eqnarray}
\label{peri8}
M(r,t)=\frac{1}{(2dt)^2}\left (1-\sqrt{1-4dtr^d}\right )^2.
\end{eqnarray}
According to Eq. (\ref{f8}), the corresponding density profile is
\begin{eqnarray}
\label{peri9}
\rho(r,t)=\frac{1}{S_d t}\left (\frac{1}{\sqrt{1-4dtr^d}}-1\right ).
\end{eqnarray}
It is represented in Fig. \ref{peripheric} at different times. Since $r\le r_{max}(t)$, we can check that the term under the square root is always positive. It vanishes for $r=r_{max}$ at the collapse time
\begin{eqnarray}
\label{peri10}
t_{coll}=\frac{1}{2d}.
\end{eqnarray}
At that time, the density becomes infinite at $r=r_{max}(t_{coll})=(1/2)^{1/d}$. On the other hand, we note that the central density $\rho(0,t)=0$ at any time. The density profile is self-similar since
\begin{eqnarray}
\label{peri11}
\rho(r,t)=\frac{1}{t}f(r t^{1/d}),
\end{eqnarray}
with the invariant profile
\begin{eqnarray}
\label{peri12}
f(x)=\frac{1}{S_d}\left (\frac{1}{\sqrt{1-4d x^d}}-1\right ).
\end{eqnarray}
This function increases and diverges when $x\rightarrow (1/4d)^{1/d}$. We must distinguish two cases:

(i) For $t<t_{coll}$, a simple calculation shows that
\begin{eqnarray}
\label{peri13}
\rho(r_{max}(t),t)=\frac{2d}{S_d}\frac{1}{1-2dt},
\end{eqnarray}
\begin{eqnarray}
\label{peri14}
M(r_{max}(t),t)=1.
\end{eqnarray}
In that case, the profile (\ref{peri8})-(\ref{peri9}) contains all the mass and there is no Dirac peak. The density at the periphery $\rho(r_{max}(t),t)$ increases monotonically with time and diverges when $t=t_{coll}$. The density profile at $t=t_{coll}$ is
\begin{eqnarray}
\label{peri15}
\rho(r,t_{coll})=\frac{2d}{S_d}\left (\frac{1}{\sqrt{1-2r^d}}-1\right ),
\end{eqnarray}
for $r<r_{max}(t_{coll})=(1/2)^{1/d}$ and $\rho(r,t)=0$ otherwise.

\begin{figure}[!h]
\begin{center}
\includegraphics[clip,scale=0.3]{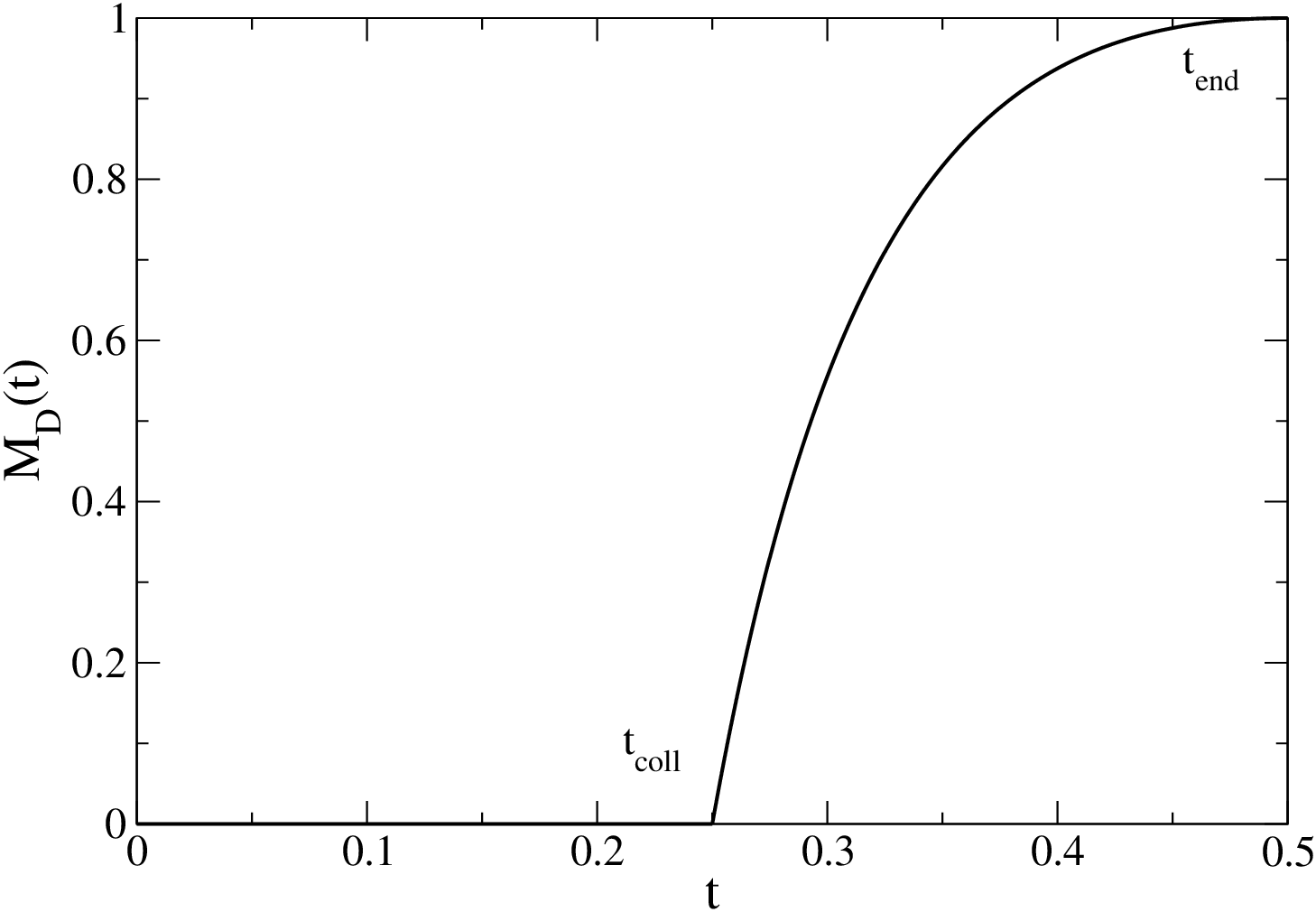}
\caption{Evolution of the mass contained in the peripheric Dirac peak in $d=2$.}
\label{diracperif}
\end{center}
\end{figure}

(ii) For $t>t_{coll}$, a simple calculation shows that
\begin{eqnarray}
\label{peri16}
\rho(r_{max}(t),t)=\frac{2}{S_d t}\frac{1-dt}{2dt-1},
\end{eqnarray}
\begin{eqnarray}
\label{peri17}
M(r_{max}(t),t)=\frac{1}{(dt)^2}(1-dt)^2.
\end{eqnarray}
In that case, the profile (\ref{peri8})-(\ref{peri9}) does {\it not} contain all the mass. A Dirac peak forms at $r=r_{max}(t)$ and captures the peripheric mass as the system shrinks. The mass contained in the Dirac peak is $M_D(t)=1-M(r_{max}(t),t)$ yielding
\begin{eqnarray}
\label{peri18}
M_D(t)=\frac{2dt-1}{(dt)^2}.
\end{eqnarray}
It obviously satisfies $M_D(t_{coll})=0$ and $M_D(t_{end})=1$ (see
Fig. \ref{diracperif}). The density at the periphery is infinite at
$t=t_{coll}$ and monotonically decreases for $t>t_{coll}$ as the Dirac
peak grows by absorbing the neighboring particles. The total
(normalized) density profile can be written
\begin{eqnarray}
\label{peri19}
\rho_{tot}({\bf r},t)=\rho({\bf r},t)+M_D(t)\frac{\delta(r-r_{max}(t))}{S_d r^{d-1}},
\end{eqnarray}
where the first term $\rho({\bf r},t)$ is the regular density profile (\ref{peri9}) and the second term corresponds to the Dirac peak. For $t\rightarrow t_{coll}$ and $r\rightarrow r_{max}(t)^{-}$, the regular density profile (\ref{peri9}) has the  self-similar form 
\begin{eqnarray}
\label{peri19b}
\rho(r,t)=\frac{1}{|t-t_{coll}|}\frac{1}{S_d\sqrt{1+\frac{2^{1/d-2}}{d}\frac{r_{max}(t)-r}{(t-t_{coll})^2}}}.
\end{eqnarray}

We can understand the onset of the Dirac peak formation in a more
qualitative manner. According to Eq. (\ref{peri3}), the positions of
the particles initially located at $a_1$ and $a_2>a_1$ coincide at the
time
\begin{eqnarray}
\label{peri20}
t=\frac{1}{d}\frac{1}{a_1^d+a_2^d},
\end{eqnarray}
and take the value
\begin{eqnarray}
\label{peri21}
r_1=r_2=\frac{a_1 a_2}{(a_1^d+a_2^d)^{1/d}}.
\end{eqnarray}
The shortest time at which this coincidence occurs corresponds to $a_1\sim a_2\sim 1$ yielding $t=t_{coll}=1/(2d)$ and $r_1=r_2=(1/2)^{1/d}$. At that time, the density becomes infinite since there is a finite mass in the interval $r_2-r_1\rightarrow 0$. Then, a peripheric Dirac peak with mass $M_D(t)$ forms at $r_{max}(t)$ and grows  by capturing the neighboring particles as the system collapses. At $t=t_{end}$, the Dirac annulus has reached the origin and all the particles are contained in a central Dirac peak at $r=0$.

{\it Remark:} According to Eq. (\ref{peri3}), we would naively conclude that the particle initially at $a$ reaches the origin $r=0$ at time $t_*(a)=1/(da^d)$. This time decreases with $a$ so that the particles that are initially close to the origin seem to take more time to collapse at $r=0$ than far away particles. In particular, for $a<1$, we find that $t_*(a)>t_{end}$ which leads to an apparent paradox. In fact, the above estimate is not correct. Indeed, a particle initially located at $a$ is captured by the peripheric Dirac peak at a time $t_c(a)$ such that $r(t_c(a))=r_{max}(t_c(a))$. This leads to $t_c(a)=(1-a^d)/\lbrack d(1-a^{2d})\rbrack$. For $a\rightarrow 1$, we find that $t_c(a)\rightarrow t_{coll}=1/(2d)$ which corresponds to the onset of the peripheric Dirac peak. We also note that $t_c(a)<t_*(a)$ for $a\le 1$, so that the particles are captured by the peripheric Dirac peak before they can reach the origin according to their pure motion (\ref{peri3}). Therefore, they are carried towards the origin by the Dirac peak whose equation of motion is given by Eq. (\ref{peri5}). This solves the apparent paradox.

\section{Repulsive interaction}
\label{sec_repu}

In the main part of the paper, we have considered the situation where the interaction between particles is attractive like gravity ($G>0$). In this Appendix, we briefly consider the case of a repulsive interaction like in a Coulombian plasma ($G<0$). In chemotaxis, when the chemical substance secreted by the biological entities acts as a {\it pheromone}, the interaction is attractive (chemoattraction) but when it acts as a {\it poison}, the interaction is repulsive (chemorepulsion). The case of a repulsive interaction can therefore have interesting physical applications.

In the deterministic case ($T=0$), the general exact solution of the problem is given by Eqs. (\ref{f9}) and (\ref{f11}) by simply making the substitution $G\rightarrow -G$. Accordingly, the dynamical evolution of a uniform sphere is given by Eqs. (\ref{h1})-(\ref{h7}) with $G$ replaced by $-G$. The sphere remains uniform during the evolution but it now expands instead of collapsing. The mass profile, the radius and the density evolve like
\begin{eqnarray}
\label{repu1}
M(r,t)=\frac{M}{R^d}\frac{r^d}{1+\frac{dGM}{\xi R^d}t},
\end{eqnarray}
\begin{eqnarray}
\label{repu2}
R(t)=R\left (1+\frac{dGM}{\xi R^d}t\right)^{1/d},
\end{eqnarray}
\begin{eqnarray}
\label{repu3}
\rho(t)=\frac{\rho(0)}{1+\frac{dGM}{\xi R^d}t}.
\end{eqnarray}
For $t\rightarrow +\infty$, we obtain the scalings $M(r,t)\sim \xi r^d/(dGt)$, $R(t)\sim (dGMt/\xi)^{1/d}$ and $\rho(t)\sim \xi/(S_d Gt)$.

For the initial parabolic profile (\ref{p1})-(\ref{p4}), the exact time dependent solution can be written
\begin{eqnarray}
\label{repu4}
M(r,t)=A a^d (1-Ba^2),
\end{eqnarray}
\begin{eqnarray}
\label{repu5}
x^d=\frac{d}{\xi} GA a^d (1-Ba^2)+\frac{a^d}{t},
\end{eqnarray}
\begin{eqnarray}
\label{repu6}
x=\frac{r}{t^{1/d}}.
\end{eqnarray}
The corresponding density profile is
\begin{eqnarray}
\label{repu7}
\rho(r,t)=\frac{\xi}{S_d Gt}\frac{1-\frac{a^2}{R^2}}{1-\frac{a^2}{R^2}+\frac{\xi}{dGAt}}.
\end{eqnarray}
The size of the system, i.e. the position of the last particle initially located at $a=R$ at time $t=0$, is
\begin{eqnarray}
\label{repu8}
r_{max}(t)=R\left (1+\frac{dGM}{\xi R^d}t\right)^{1/d}.
\end{eqnarray}
We note that this expression coincides with Eq. (\ref{repu2}) for all times. The central density $(r=a=x=0)$ decreases like
\begin{eqnarray}
\label{repu9}
\rho(0,t)=\frac{\xi}{S_d Gt}\frac{1}{1+\frac{\xi}{dGAt}}.
\end{eqnarray}
For $t\rightarrow +\infty$, the system behaves like a homogeneous sphere since Eqs. (\ref{repu4})-(\ref{repu9}) become equivalent to Eqs. (\ref{repu1})-(\ref{repu3}) in this limit. Considering the general equations of the problem, we note that this property is valid for an arbitrary initial profile. For $t\rightarrow 0$, we find that
\begin{eqnarray}
\label{repu10}
\rho(r,t)\simeq \rho(r,0)-\frac{d^2(d+2)^2}{4 S_d}\frac{GM^2}{\xi R^{2d}}\nonumber\\
\times \left\lbrack 1-\frac{2(d+1)}{d}\frac{r^2}{R^2}+\frac{d+4}{d+2}\frac{r^4}{R^4}\right\rbrack t.
\end{eqnarray}
The corresponding mass profile is
\begin{eqnarray}
\label{repu11}
M(r,t)\simeq M(r,0)-\frac{d(d+2)^2}{4}\frac{GM^2}{\xi R^{2d}}r^d\nonumber\\
\times \left\lbrack 1-\frac{2(d+1)}{d+2}\frac{r^2}{R^2}+\frac{d}{d+2}\frac{r^4}{R^4}\right\rbrack t.
\end{eqnarray}
These expressions are valid for $r\le r_{max}(t)$ with
\begin{eqnarray}
\label{repu12}
r_{max}(t)\simeq R\left (1+\frac{GMt}{\xi R^d}\right ).
\end{eqnarray}
For $t\rightarrow 0$, the central density evolves like
\begin{eqnarray}
\label{repu13}
\rho(0,t)\simeq \frac{dA}{S_d}\left (1-\frac{dGAt}{\xi}\right ).
\end{eqnarray}

\begin{figure}[!h]
\begin{center}
\includegraphics[clip,scale=0.3]{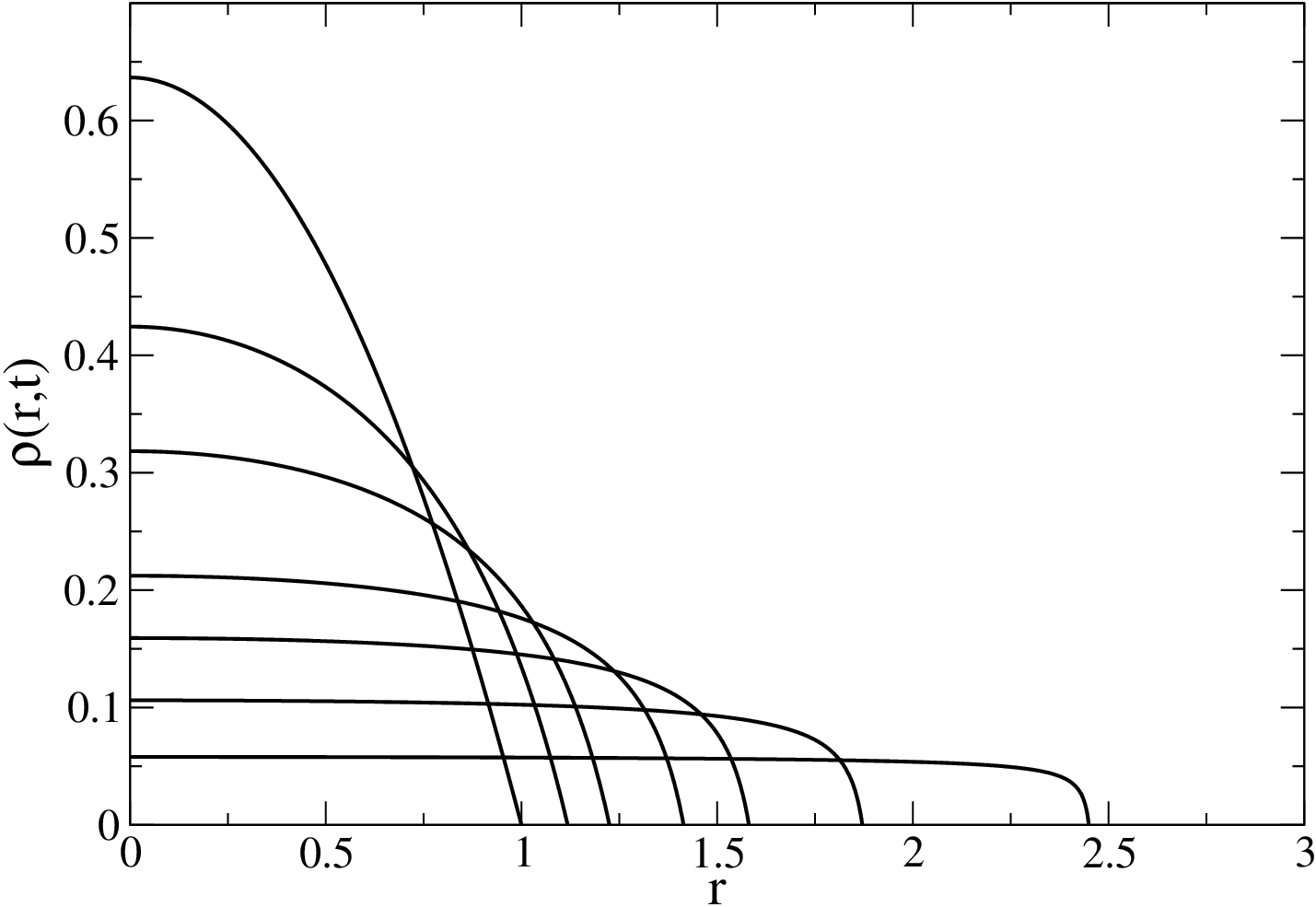}
\caption{Evolution of the density profile $\rho(r,t)$ in $d=2$ for $t=0, 0.5, 1,2,3,5,10$. The initial density profile is parabolic but it asymptotically tends towards a step function (uniform disk). }
\label{repulsive}
\end{center}
\end{figure}

In $d=2$, Eq. (\ref{repu5}) can be solved explicitly to obtain $a(x,t)$. We get
\begin{eqnarray}
\label{repu14}
a^2=\frac{1}{2B}\left\lbrack 1+\frac{\xi}{2GAt}-\sqrt{\left (1+\frac{\xi}{2GAt}\right )^2-\frac{2\xi Bx^2}{GA}}\right \rbrack.\nonumber\\
\end{eqnarray}
Substituting this expression in Eqs. (\ref{repu4}) and (\ref{repu7}), we obtain an explicit solution for $M(r,t)$ and $\rho(r,t)$. To simplify the notations, we introduce dimensionless variables such that $M=R=1$, $A=2$, $B=1/2$, $\xi/G=4$ and $S_2=2\pi$ like in Sec. \ref{sec_illustration}. Equations (\ref{repu7}) and (\ref{repu14}) can be combined to give
\begin{eqnarray}
\label{repu15}
\rho(r,t)=\frac{2}{\pi t}\left (1-\frac{1}{\sqrt{(1+t)^2-2r^2t}}\right ),
\end{eqnarray}
for $r\le r_{max}(t)$ with
\begin{eqnarray}
\label{repu16}
r_{max}(t)=\left (1+\frac{t}{2}\right )^{1/2}.
\end{eqnarray}
For $r\ge r_{max}(t)$, we have $\rho(r,t)=0$ as usual. The density profile is plotted in Fig. \ref{repulsive} for different times. For $t=0$, the profile is parabolic and for $t\rightarrow +\infty$, it tends to a Heaviside function (uniform disk).  The central density decreases like
\begin{eqnarray}
\label{repu17}
\rho(0,t)=\frac{2}{\pi}\frac{1}{1+t}.
\end{eqnarray}

In the repulsive case, making the substitution $G\rightarrow -G$ in Eqs. (\ref{vir2})-(\ref{vir3}), the virial theorem in $d=2$ dimensions reads
\begin{eqnarray}
\label{repu18}
\frac{1}{4}\xi\frac{dI}{dt}=N\left (k_B T+\frac{GMm}{4}\right ).
\end{eqnarray}
It can be integrated into
\begin{eqnarray}
\label{repu19}
I(t)=\frac{4N}{\xi}\left (k_B T+\frac{GMm}{4}\right )t+I(0).
\end{eqnarray}
For $T=0$, we obtain
\begin{eqnarray}
\label{repu20}
I(t)=\frac{GM^2}{\xi}t+I(0).
\end{eqnarray}
Introducing the mean square displacement $\langle r^2\rangle=I(t)/M$, Eq. (\ref{repu19}) shows that the motion of a particle is diffusive with an effective diffusion coefficient
\begin{eqnarray}
\label{repu21}
D(T)=\frac{k_BT}{\xi m}+\frac{GM}{4\xi},
\end{eqnarray}
that is independent on the initial condition. For $T=0$, we obtain $D_0=GM/(4\xi)$. For the uniform sphere in $d$ dimension, using Eqs. (\ref{repu1})-(\ref{repu3}), we get
\begin{eqnarray}
\label{repu22}
I(t)=\frac{d}{d+2}MR^2\left (1+\frac{dGM}{\xi R^d}t\right )^{2/d}=\frac{d}{d+2}MR(t)^2.\nonumber\\
\end{eqnarray}
In $d=2$, this expression reduces to Eq. (\ref{repu20}) as it should. For a parabolic initial condition, introducing the dimensionless variables defined previously and using Eq. (\ref{repu15}), we find after some calculations that
\begin{eqnarray}
\label{repu23}
I(t)=\langle r^2\rangle=\frac{t}{4}+\frac{1}{3}.
\end{eqnarray} 
This expression coincides with Eq. (\ref{repu20}), as it should.

\end{document}